\documentclass[onecolumn,secnumarabic,amssymb,nobibnotes,aps,pre]{revtex4} 

\usepackage{setspace}

\usepackage{amsmath}
\usepackage{array}
\usepackage[centerlast,small,bf]{caption}
\usepackage[bf,FIGTOPCAP,large,nooneline]{subfigure}
\usepackage{float}
\usepackage{graphicx}
\usepackage{dcolumn}
\usepackage{bm}

\begin{document}

\title{The effect of the filament-obstacle interaction on the force-velocity relation of a growing biopolymer}

\author{F.~Motahari}
\author{A.~E.~Carlsson}
\email{aec@wustl.edu}%
\affiliation{Department of Physics and Center for Engineering Mechanobiology, Washington University, St.
Louis, Missouri 63130, USA}

\date{\today}

\pagestyle{myheadings} \markright{Biopolymer Force-Velocity Relation}

\begin{abstract}
We investigate the effect of filament-obstacle interactions on the force-velocity relation of
growing biopolymers, via calculations explicitly treating obstacle diffusion and stochastic
addition and subtraction of subunits. We first show that the subunit on- and off-rates satisfy
a rigorous thermodynamic relationship determined by the filament-obstacle
interaction potential. Both the on- and off-rates depend not only on the average force
on the obstacle, but also on the shape of the potential on the nanometer length scale. 
Basing obstacle-induced reduction of the on-rate
entirely on the force overestimates the stall force when there are fluctuations
in the force exerted on a filament tip.  We then perform
simulations and analytic calculations using the thermodynamic relationship. We find,
consistent with expectations from general thermodynamic relations, that the ``Brownian-Ratchet"
model is an upper bound to the growth velocity and that for purely repulsive potentials
the growth velocity is essentially that predicted by the Brownian-Ratchet model. 
For shallow potential wells of depth $\sim 5k_BT$, which might correspond to transient 
filament-membrane attachments, the zero-force velocity is a substantial fraction of the free-filament velocity.
In this case, the growth velocity can depend strongly on the obstacle diffusion coefficient 
even when the dimensionless diffusion coefficient is large. 
The velocity also drops more rapidly than predicted by the Brownian-ratchet model,
in some cases by as much as a factor of 50 at an opposing force of $1$ pN.  
For deep potential wells, as might result from
strong filament-membrane links,  both the on- and off-rates are reduced significantly, 
slowing polymerization. Such potentials can sustain pulling forces while polymerizing, but only
if the attractive well has a ``shelf" comparable to or greater than the monomer size. 
For such potentials, the slowing of polymerization by external force
is almost linear. 
\end{abstract}


\maketitle

\section{Introduction}

Polymerization of biopolymers such as actin filaments provides force to drive both protrusion and invagination of the cell membrane. 
This process is modulated by the interaction between the filament tip and the membrane, which varies considerably
between different cellular phenomena. In lamellipodia, filaments are likely weakly bound to the membrane by their interaction with, for example, WASP-family actin-nucleating proteins \citep{bieling2017wh2}. On the other hand, in filopodia actin filaments are nucleated by formins, which can bind actin filaments strongly \citep{yu2017mdia1}. Furthermore,
 in processes that require strong pulling forces to bend the membrane, such as endocytosis in yeast, 
some actin filaments must be strongly bound to the endocytic site. In this process, the binding is believed to result from the protein Sla2, which has both actin-binding domain and a domain that links it to the membrane. \citep{kaksonen2003pathway}. Recent super-resolution experiments have shown that WASP (Las17 in yeast) forms a ring around a Sla2 dot \citep{mund2018systematic}, and pulling forces are probably concentrated in the dot \citep{tweten2017actin, wang2016actin}. Similarly, microtubule interactions with the cell membrane are mediated by a range of proteins, which
may lead to a variety of effective interaction potentials \citep{waterman1995membrane,perez1999clip}.
In all of these cases, it is important to understand how the interaction between a growing  filament and the membrane plus associated proteins affects the polymerization rate. The filament-membrane interaction may be optimized for different criteria, such as polymerization velocity where migration speed is crucial, or stability where strong pulling forces are required.

Most previous calculations of force generation by polymerization have used hard wall repulsive potentials acting between the filament tip and the obstacle. The classic Brownian ratchet (BR) model \citep{peskin1993cellular} used such a potential to treat polymerization in the presence of a diffusing obstacle, assuming that the monomer on-rate increases suddenly from zero to its free-filament value at a certain distance from the membrane. Analysis of this model showed that in the limit of fast obstacle diffusion, the hard-wall potential gives rise to a growth velocity that decays exponentially with opposing force:
\begin{equation}
v_{growth} = \delta [k_{on}^0  \exp{( -F \delta/k_BT )}  - k_{off}^{0}],
\label{BR}
\end{equation}
where $\delta$ is the step size per added subunit, and $k_{on,off}^0$ are the free-filament on- and off-rates.

Calculations based on this analysis can qualitatively explain the mechanism behind the generation of pushing forces by actin polymerization in several types of cellular protrusions, such as filopodial and lamellipodial protrusions \citep{mogilner2002regulation, atilgan2005morphology, mogilner2005physics, rubinstein2005multiscale, maree2006polarization, ditlev2009open, joanny2009active, khamviwath2013continuum, camley2017crawling, zimmermann2012actin}. The basic features of intracellular motion of pathogens such as Listeria, which is driven by actin polymerization, can also be explained by BR-type models.
	I should also see if the geometry affects the efficiency (eout over ein) more than it affects eout. 

However, the true interaction is continuously varying and the force may have both attractive and repulsive components. It is not known how these variations affect the force-velocity relation. For example, a potential with a deep well might trap the growing end of the filament near the obstacle, and thus slow polymerization. On the other hand, a smoother potential might speed polymerization. A small number of calculations have treated such smoothly varying potentials. Calculations in 2D using an explicitly moving obstacle with explicitly diffusing monomers to treat the growth rate of a single actin filament, interacting with an obstacle via a steeply increasing force field, found that the velocity decays more rapidly than the BR prediction \citep{carlsson2000force}. This effect was attributed to a diffusion barrier in which monomers had to traverse a tunnel-like region to reach their binding site at the end of the filament. Such effects are expected to be smaller in 3D. Calculations using a range of force fields to treat 3D polymerization \citep{burroughs2005three,burroughs2006growth}, including
filament bending, found large acceleration of growth by a soft obstacle when the obstacle diffusion coefficient was small. 
Refs. \citep{mogilner2003force} and \citep{banigan2013control} considered actin filaments strongly bound to the obstacle, but assumed they did not grow. 
Studies of filament growth while clamped to a motile obstacle via a deep potential energy well suggested that the filament-obstacle attachment is the controlling factor for the elongation rate \citep{dickinson2002clamped,dickinson2004force}.
A later model considered filaments attached to an obstacle with a double-well potential \citep{zhu2006growth}, explicitly
treating diffusive motion of the obstacle.  It was found that a filament can push the obstacle and grow with a speed of about half of the free filament speed and thus progressively polymerize, if the potential is sufficiently deep. 

Among these preceding studies, there is no systematic exploration of a broad range of possible force fields, to establish how the 
force field influences the force-velocity relation. 
In fact, the methodology for performing calculations with smoothly
varying filament-obstacle interactions is not well established. One must choose spatial dependences for both 
the interaction energy between the obstacle and the filament tip, 
and the polymerization rate parameters. In previous work these dependences have often been chosen independently. 

Here we establish a thermodynamic relationship between the spatial dependence of 
the polymerization-rate parameters and that of the interaction energy, simplifying the construction of appropriate force-generation models. 
This relationship applies to a filament growing against a diffusing obstacle whose motion is treated explicitly, with a smoothly 
varying filament-obstacle interaction.  
Performing calculations without this relationship can lead to an incorrect stall force.
In our implementation, either the on-rate or off-rate is reduced, depending on 
the form of the interaction potential; neither is increased. 
We then perform a systematic set of simulations for a broad range of possible filament-obstacle 
interaction potentials, treating polymerization and depolymerization as well as obstacle motion stochastically. 
We find that monotonically decaying repulsive potentials lead to force-velocity
relations very similar to the BR prediction, as expected from general thermodynamic principles.  
Weak attractive potentials both reduce the zero-force velocity, and 
lead to a decay that is more rapid than the Brownian-ratchet prediction. 
Deep and narrow potentials lead to slow polymerization at all force values. 
We find that attached filaments stop growing if the potential well is 
deep enough to sustain pulling forces greater than about 1pN, unless the potential well has a shelf
comparable to the monomer size.  Potentials with such a shelf
have fairly rapid polymerization at zero force and have a force-velocity relation that decays almost linearly.  

Although the model is highly simplified, multiscale calculations such as those of Ref. \citep{mogilner2002regulation, mogilner2005physics,ryan2017cell,zimmermann2012actin,craig2012membrane,adler2013closing,carlsson2014force,barnhart2017adhesion} have demonstrated the utility of simple, but approximate results for the force-velocity relation in calculating the properties of cells and processes inside cells. The general understanding gained from the present studies will enhance the physical relevance of such multiscale calculations. 

\section{Model}

\subsection{Filament-Obstacle Interaction}

\begin{figure}[H]
\centering
\includegraphics[width=0.5\textwidth]{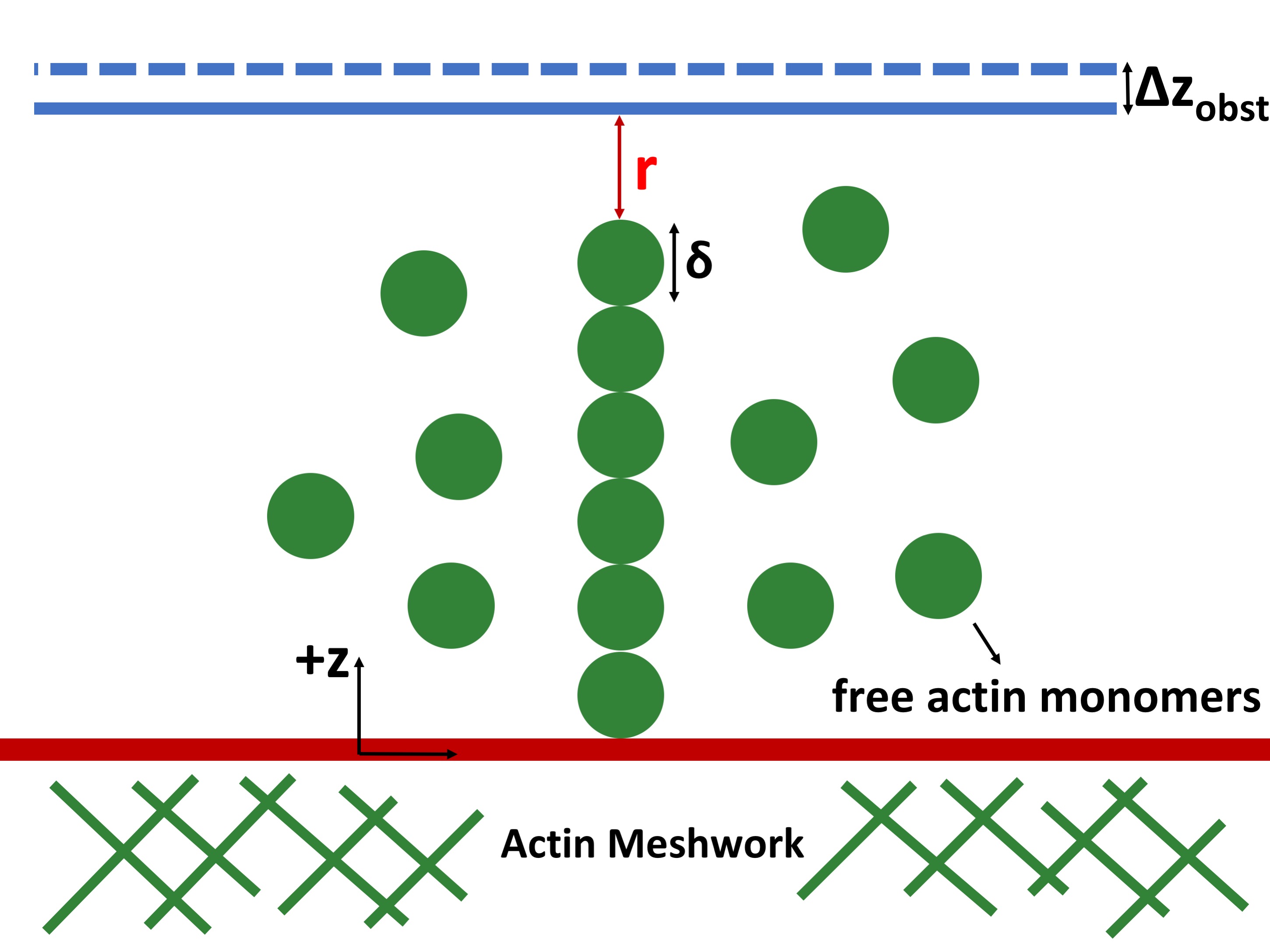}
\caption{Schematic of model applied to an actin filament: $r$ is the distance between the filament tip and the obstacle, $\delta$ is the actin monomer size = 2.7nm, and $\Delta z_{obst}$ is the obstacle position fluctuation in the +z direction during a given time step. We treat the filament's base as being solidly anchored.}
\label {model_fig}
\end{figure}

\begin{figure}[h]
\centering
\subfigure[]{
\label{Pot}
\includegraphics[width=0.45\textwidth]{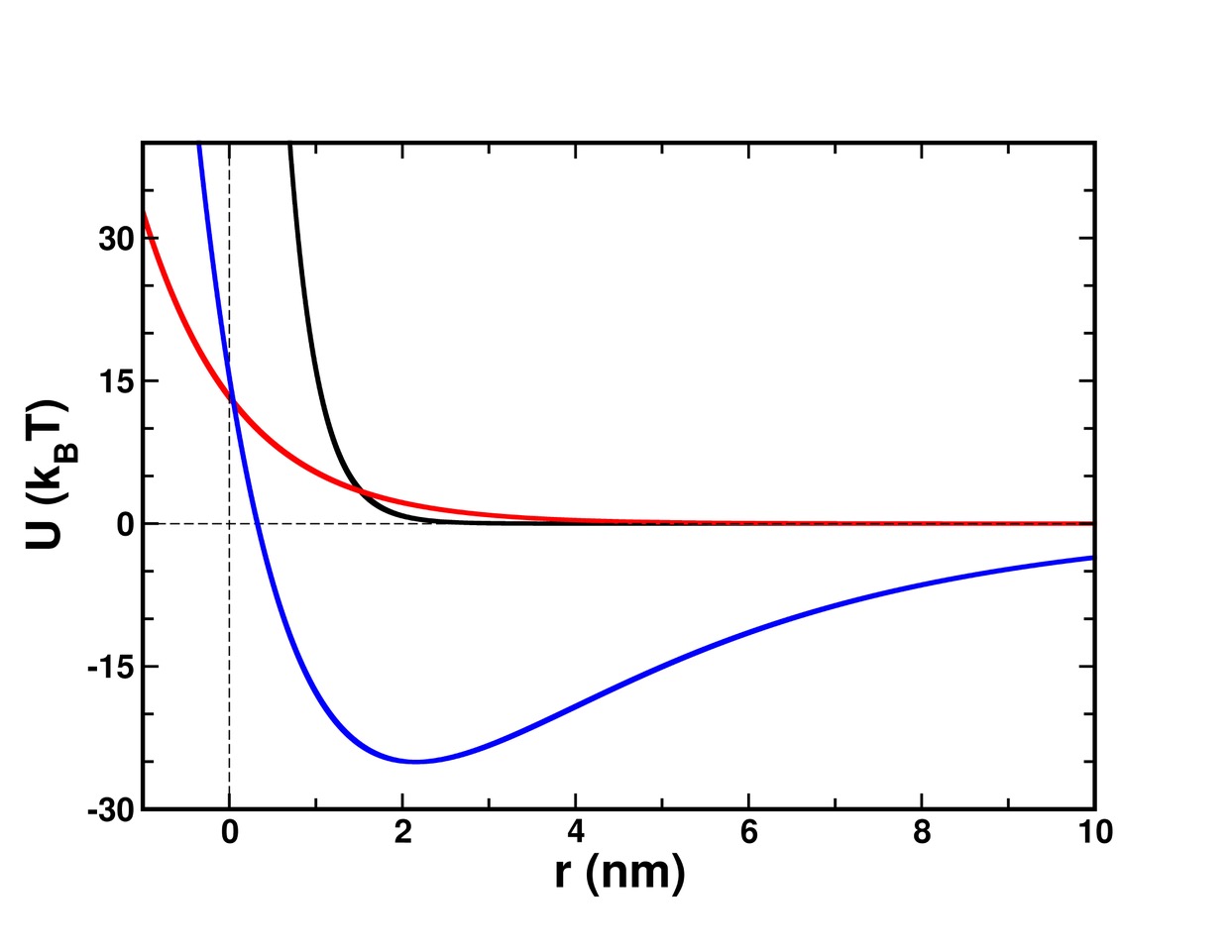}}
\qquad
\subfigure[]{
\label{Spikes}
\includegraphics[width=0.45\textwidth]{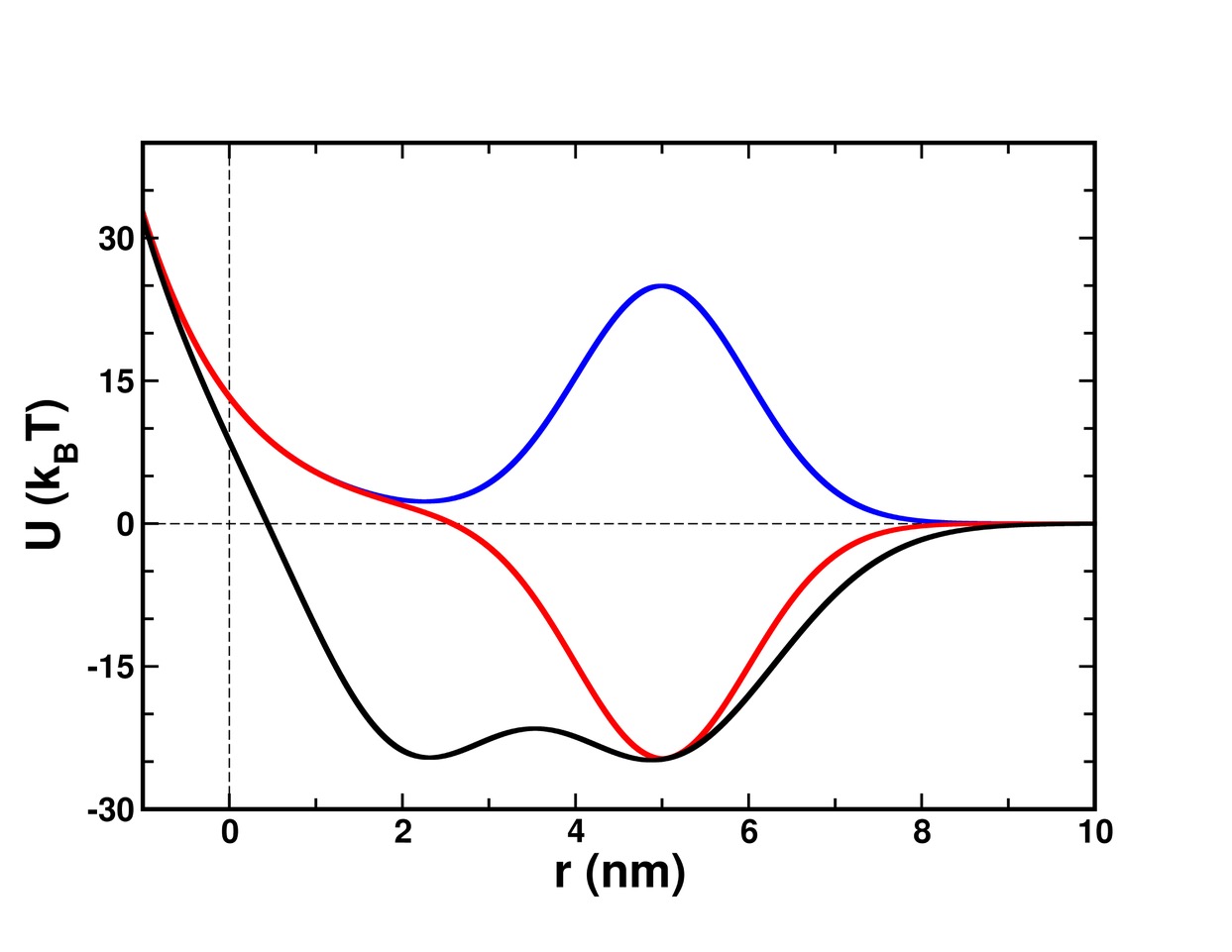}}
\caption{Filament-obstacle interaction potentials.  \textbf{(a)} Potentials from Eq.~\ref{U1}. Black curve represents a hard obstacle, with A=1339\,$pN$$\cdot nm$, B=0, and $\kappa_1$=3\,$nm^{-1}$. Red curve corresponds to a soft obstacle with A=54.7\,$pN$$\cdot nm$, B=0, and $\kappa_1$=0.9\,$nm^{-1}$. Blue curve shows a simple-well potential with a depth of 25$k_BT$, with parameter values A=358\,$pN$$\cdot nm$, B=295\,$pN$$\cdot nm$, $\kappa_1$=0.9\,$nm^{-1}$, and $\kappa_2$=0.3\,$nm^{-1}$. \textbf{(b)} Potentials from Eq. \ref{U2}, containing one or more Gaussian spikes. Blue curve has a positive Gaussian spike described by parameters A=54.7\,$pN$$\cdot nm$, D=0, $\kappa_3$ = 0.707\,$nm^{-1}$, $r_{1}$ = 5\,nm, and C = -104\,$pN$$\cdot nm$; red curve differs from this one by having  C = 104\,$pN$$\cdot nm$. Black curve is a double-well potential with A=54.7\,$pN$$\cdot nm$, C=100\,$pN$$\cdot nm$, D=100\,$pN$$\cdot nm$, $\kappa_3$ = 0.643\,$nm^{-1}$, $\kappa_4$ = 0.544\,$nm^{-1}$, $r_{1}$ = 2\,nm, and $r_{2}$ = 5\,nm.}
\label{Potentials}
\end{figure}

The model (Figure~\ref{model_fig}) treats the stochastic polymerization of a biopolymer, exerting force on an explicitly moving, flat, penetrable obstacle. We envisage the base of the filament as
being rigidly anchored. Actin filaments, for example, could be anchored in a
crosslinked actin meshwork. For conceptual simplicity the results presented here are for
a filament growing perpendicular to the obstacle without bending fluctuations; results for a filament growing at an oblique angle, including membrane fluctuations, are described in the Appendices.
We treat a range of filament-obstacle interactions given by smooth potential functions having the form  
\begin{equation} \label{U1}
U(r) = Ae^{-\kappa_1r} - Be^{-\kappa_2r}
\end{equation}
or
\begin{equation} \label{U2}
U(r) = Ae^{-\kappa_1r} - Ce^{-[\kappa_3(r-r_{1})]^2} - De^{-[\kappa_4(r-r_{2})]^2}
\end{equation}
Here, A, B, C, D, $\kappa_1$, $\kappa_2$, $\kappa_3$, $\kappa_4$, $r_{1}$, and $r_{2}$ are constants, 
and $r$ is the variable gap between the tip of the filament and the obstacle.  
Filament-obstacle binding can be naturally included in the force field via the $B$ term in  Eq.~\ref{U1}; we denote
such a potential a ``simple well".  A non-zero C in Eq.~\ref{U2} also adds a Gaussian ``spike" to the potential at $r$ = $r_{1}$ which can be either attractive or repulsive. Choosing both C and D to be positive in Eq.~\ref{U2} generates a double-well potential, as shown in Figure~\ref{Potentials}(b). The depth of the wells in the potentials represents the energy of binding filaments to membrane-bound proteins. 
The corresponding forces that the filament exerts on the obstacle are
\begin{equation} \label{F1}
\left.\begin{aligned}
F(r) = -\frac{dU}{dr} = A\kappa_1e^{-\kappa_1r} - B\kappa_2e^{-\kappa_2r}
\end{aligned}\right.
\end{equation}
and
\begin{equation} \label{F2}
\left.\begin{aligned}
F(r) = A\kappa_1e^{-\kappa_1r} - 2C{\kappa_3}^2(r-r_{1})e^{-[\kappa_3(r-r_{1})]^2} - 2D{\kappa_4}^2(r-r_{2})e^{-[\kappa_4(r-r_{2})]^2},
\end{aligned}\right.
\end{equation}\\
respectively. We will denote a filament having only repulsive potential terms a  ``pusher". A filament with an interaction potential containing a well or a spike is denoted a ``puller" (although it can also exert a pushing force).

\subsection{Obstacle Dynamics}

The obstacle position is stepped forward in time according to biased Brownian motion driven by $F(r)$ and thermal fluctuations. We use a ``filament-centric" approach in our simulations, in which the filament base is assumed to be stationary and the obstacle moves. Cases where the filament base is moving can be handled by a simple coordinate transformation, in which obstacle motion is the inverse of the base motion and the diffusion coefficient of the obstacle is replaced by that of the entity to which the filament is anchored. At each time step, the diffusive motion of the obstacle is calculated by the discrete form of the Langevin equation (Ref.~\citep{doi1988theory}, Chap.~3):
\begin{equation} 
\label{z_memb}
\Delta z_{obst} = \alpha\sqrt{24\Delta t}\sqrt{D_{obst}} + \frac{D_{obst}}{k_BT}\Delta t [ F(r) + F_{load} ]
\end{equation}
where $D_{obst}$ is the obstacle diffusion constant, $F_{load}$ is the external force applied on the obstacle, $\Delta t$ is the time step, and $\alpha$ is a random number uniformly distributed between $-\frac{1}{2}$ and $\frac{1}{2}$, so that $<\alpha^2> = \frac{1}{12}$. Consecutive time steps are uncorrelated.

\section{Results}

\subsection{Thermodynamic Relation between $k_{on} (r)$ and $k_{off} (r)$}

It is physically clear that at least $k_{on}$ must be modified as a filament tip approaches an obstacle, because there is less room available for new subunits to add. Some previous models \citep{mogilner1999polymerization, van2000stall, son2005monte, krawczyk2011stall} treated hard-wall potentials and considered ${\bar k}_{on}$, the addition rate averaged over a time long in comparison with the time scale of filament-tip fluctuations. They argued that when the distance between the filament tip and the closest obstacle position to the filament tip in a multi-filament simulation is less then the monomer length increment $\delta$, ${\bar k}_{on}$ is reduced by a factor of $\exp{[-F(\delta-r)/k_BT]}$ relative to the free-filament value, where $F$ is the time-averaged force exerted on the filament tip, and $r$ is the distance between filament tip and the obstacle. This relationship holds when the force required to bend a filament tip is fairly constant over the size of a subunit. Here we show that a more complex relationship holds when the filament-obstacle interaction varies strongly over distances on the order of the subunit size. In such cases, it is necessary to include obstacle motion explicitly in the
calculations. Then  one uses rates $k_{on}(r)$ and $k_{off}(r)$ that refer to polymerization and depolymerization events occurring at a given filament-tip and obstacle position, rather than time-averaged rates.

We consider polymerization of filaments in the absence of nonequilibrium effects such as hydrolysis of ATP to ADP in actin. In this case, the stall force must be independent of the form of the interaction potential $U(r)$ between the filament tip and the obstacle. This follows from the thermodynamic arguments of Ref.~\citep{hill2012linear}: At the stall force, changes in chemical free energy resulting from polymerization precisely balance changes in mechanical energy, relating the stall force to the polymerization free energy per subunit. Then the combined system of obstacle and filament can be described by a free energy function, containing a mechanical term $F_{ext}z$ where $F_{ext}$ is the external force (measured in the direction opposite to filament growth) acting on an obstacle with coordinate $z$, a chemical term $N \Delta G$ where $N$ 
is the number of subunits in the filament and $\Delta G$ is the chemical free-energy increment per added subunit, and $U(r)$. Defining $r$ to be the obstacle-tip distance, $z = r + N \delta$. Then the total free energy as a function of $N$ and $r$ is
\begin{equation}
G_{tot} = N \Delta G + N F_{ext} \delta + F_{ext} r + U(r).
\label{Gtot}
\end{equation} 
Here the free energy is defined on a time scale shorter than the time scale of monomer addition and obstacle motion but still long enough that the free energy of an actin monomer interacting with water molecules is well defined. 

The stall force $F_{stall}$ is defined by $G_{tot}$ being independent of $N$ at a fixed value of $r$, so that 
\begin{equation}
F_{stall}= -\Delta G/\delta,
\label{fstall}
\end{equation}
as is well known \citep{peskin1993cellular, hill2012linear}. This result implies \citep{hill2012linear} that
\begin{equation}
F_{stall}= (k_BT/\delta ) \ln{(k^0_{on}/k^0_{off})},
\label{fstall_formula}
\end{equation}
where $k^0_{on}$ and $k^0_{off}$ are rates for a free filament not interacting with an obstacle. Note that $k^0_{on}$ is the on-\textit{rate} (having units of $s^{-1}$), which is the product of the on-rate constant with the free-actin monomer concentration.\\

\begin{figure}
\centering
\includegraphics[width=0.5\textwidth]{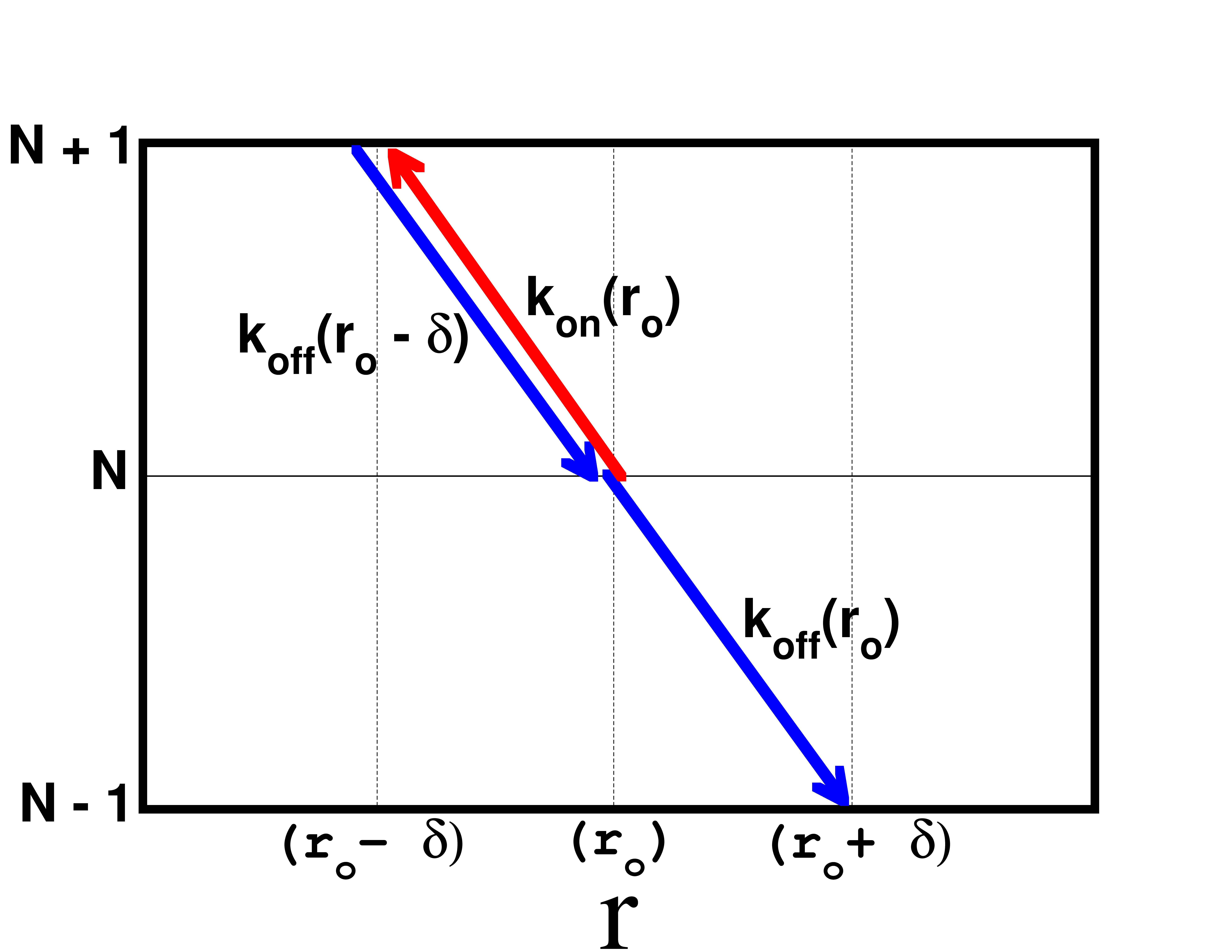}
\caption{Dynamics of obstacle motion and actin filament polymerization and depolymerization. Red arrow corresponds to the polymerization rate at the point where the filament-obstacle gap is $r_o$, and the blue arrows correspond to the depolymerization rates at $r_o$ and $r_o - \delta$.}
\label{polyscheme}
\end{figure}

Now consider the dynamics of the filament-obstacle system at the stall force. The system is in equilibrium and thus obeys detailed balance (recall that ATP hydrolysis is neglected). Since the system is at the stall force, the free energy
\begin{equation}
G_{tot} = F_{stall}\cdot r + U(r),
\label{gtot}
\end{equation} 
is independent of $N$.  The dynamic processes in the system  are i) Brownian motion of the obstacle, ii) polymerization, and iii) depolymerization.  
In Figure\,\ref{polyscheme}, Brownian motion leads to infinitesimal steps in $r$, while polymerization leads to jumps $r \rightarrow r - \delta$ and depolymerization leads to jumps $r \rightarrow r + \delta$. Because the system is in equilibrium, the probability distribution $P(r;N)$ satisfies the Boltzmann relation. In particular, referring to Figure~\ref{polyscheme} and ignoring normalization of $P$, 
\begin{eqnarray}
P(r;N) & = & \exp{\{ -[U(r)+F_{stall}\cdot r]/k_BT \} } \\
P(r-\delta;N+1) & = & \exp{\{ -[U(r-\delta)+F_{stall}\cdot(r-\delta)]/k_BT \}} \nonumber 
\end{eqnarray}
Detailed balance between the states $(r;N)$ and $(r-\delta;N+1)$, which holds for this equilibrium system, implies that
\begin{equation}
k_{on}(r) P(r;N) = k_{off}(r-\delta)P(r-\delta;N+1)
\end{equation}
Therefore, using Eq.~\ref{fstall_formula},
\begin{eqnarray}
\hspace*{-1.5cm} \frac{k_{on}(r)}{k_{off}(r-\delta )} & =&  \exp{\{ -[U(r-\delta )-U(r)]/k_BT \}} \cdot \exp{(F_{stall}\cdot \delta /k_BT)} \nonumber \\
& = &  \exp{\{ -[U(r-\delta )-U(r)]/k_BT \}} \cdot  \frac{k^0_{on}}{k^0_{off}}
\label{rigorous}
\end{eqnarray}

This relationship guarantees that the correct stall force is obtained. 
It generalizes the well-known result \citep{hill2012linear} for the average rates that 
\begin{equation}
{\bar k}_{on}/{\bar k}_{off}= \exp{(-F\delta/k_BT)}{k}_{on}^0/{k}_{off}^0.
\label{hill}
\end{equation}
Eq. \ref{hill} follows from Eq. \ref{rigorous} in the limit $D_{obst} \rightarrow \infty$. To see this,  
note that from  Eq. \ref{rigorous} the rate constants have the form 
\begin{eqnarray}
k_{off}(r)& =& k_{off}^0 f(r) \nonumber \\
k_{on}(r)&=& k_{on}^0 \exp{\{ -[U(r-\delta )-U(r)]/k_BT \}} f(r-\delta)
\label{kform}
\end{eqnarray}
where $f$ is a function of $r$.
When $D_{obst} \rightarrow \infty$, the obstacle position distribution $P(r)$ has its thermal equilibrium
form $P(r) = \exp{\{ -[U(r)+F r]/k_BT}\}/Z$, where $Z=\int_{-\infty}^{\infty} \exp{ \{ -[U(r)+F r]/k_BT} \}dr$.
Then 
\begin{eqnarray}
{\bar k}_{on} & = & (k_{on}^0/Z) \int_{-\infty}^{\infty} \exp{\{ -[U(r-\delta )-U(r)]/k_BT \}} \exp{\{ -[U(r)+F r]/k_BT \}}f(r -\delta)dr \nonumber \\
& = & (k_{on}^0/Z) \int_{-\infty}^{\infty} \exp{\{ -[U(r-\delta )]/k_BT \}} \exp{[ -F r/k_BT ]}f(r-\delta)dr \nonumber \\
& = & (k_{on}^0/Z) \exp{( -F\delta/k_BT )} \int_{-\infty}^{\infty} \exp{[ -U(r)/k_BT ]} \exp{[ -F r/k_BT ]}f(r)dr
\label{konbar}
\end{eqnarray}
while
\begin{eqnarray}
{\bar k}_{off} & = &  (k_{off}^0/Z) \int_{-\infty}^{\infty} \exp{[ -U(r)/k_BT ]} \exp{[ -F r/k_BT ]}f(r)dr   \\
& = & {\bar k}_{on} [ \exp{( F\delta/k_BT )k_{off}^0/k_{on}^0}],
\label{koffbar}
\end{eqnarray}
implying that Eq. \ref{hill} holds. When obstacle diffusion is not rapid, Eq. \ref{hill} will not necessarily hold because
energy is dissipated by the obstacle drag, which is inversely proportional to $D_{obst}$ according to the
Einstein relation. This is not accounted for in the thermodynamic analysis. The stall force,
however, is unaffected by obstacle drag because the obstacle is stationary on average. 

Eq. \ref{rigorous} correctly implies that the force-velocity relation is independent of the choice of zero
for evaluating $r$. For example, adding a constant shift $\Delta r$ to $r$ would cause the typical positions sampled by
the filament tip to move out a distance $\Delta r$ from the obstacle, i. e. remaining close to the minimum of
$U(r)$ if the potential has a deep well. Then the values sampled by the factor $\exp{\{ -[U(r-\delta )-U(r)]/k_BT\}}$
will also remain the same, corresponding to the energy difference between a point at the minimum and one
shifted in by $\Delta r$ from the minimum. This also implies that for the repulsive potential, the 
force-velocity relation is independent of the prefactor $A$; changes in the prefactor can be accounted
for by changing the zero of the $r$-coordinate, which does not affect the force-velocity relation.
This is confirmed by our numerical simulations below.

The derivation above applies to a single-stranded filament growing perpendicular to the obstacle. 
However, Eq. \ref{rigorous} holds for a broader range of models. If the filament grows at an angle of $\theta$
relative to the obstacle, then the schematic of Figure \ref{polyscheme} holds provided that $\delta$ is
replaced by $\delta \cos{(\theta )}$, the step size per added subunit. Similarly, for multistranded filament
growth Figure \ref{polyscheme} applies provided that a new subunit can add only at a unique specified site (typically 
next to the preceding one); then $\delta$ is again the step size per added subunit.  Thus Eq. \ref{rigorous}
holds for both these cases. 
It also holds when filament bending degrees of freedom are included, and for systems of many filaments. 
If, for example, one describes the bending of a single filament by angle $\theta$, then Figure \ref{polyscheme} applies to
transitions occurring at a given value of $\theta$. Because detailed balance must hold for all transitions
in a system at equilibrium, Eq. \ref{rigorous} will still hold. 
In systems of many filaments, Figure \ref{polyscheme} would apply to a single filament,
and again transitions involving just that filament must satisfy detailed balance at the stall force. 

Eq. \ref{rigorous} does not uniquely determine $k_{on} (r)$ and $k_{off}(r)$. In our simulations, we 
make a minimal assumption, that has frequently been used for the rates ${\bar k}_{on}$ and ${\bar k}_{off}$, 
by preventing both rates from exceeding the free filament on and off rates:
\begin{eqnarray}\label{k_on_eq}
k_{on}(r)  & = & k_{on}^{0}  \exp{\{ -[U(r-\delta )-U(r)]/k_BT \}} \quad \quad \textrm{if} \quad U(r-\delta) > U(r) \nonumber \\
k_{on}(r) & = & k_{on}^{0} \qquad \qquad \qquad \qquad \qquad \qquad \qquad \qquad \, \textrm{if} \quad U(r-\delta) < U(r)
\label{correctionkon}
\end{eqnarray}
\begin{eqnarray}\label{k_off_eq}
k_{off}(r)  & = & k_{off}^{0}  \exp{\{ -[U(r+\delta )-U(r)]/k_BT \}} \quad \quad \textrm{if} \quad U(r+\delta) > U(r) \nonumber \\
k_{off}(r) & = & k_{off}^{0} \qquad \qquad \qquad \qquad \qquad \qquad \qquad \qquad \, \textrm{if} \quad U(r+\delta) < U(r)
\label{correctionkoff}
\end{eqnarray}
Thus if $U(r)$ is monotonically repulsive, there is no correction to $k_{off}$ in Eq. \ref{correctionkoff}.
This assumption has been made in most previous calculations in the literature. We do not have strong arguments
justifying the assumption, but have decided to make it here in order to avoid investigating an unwieldy
set of possibilities. 

In the limit of a hard obstacle, where $U(r - \delta )$ jumps suddenly from $0$ to $\infty$  when $r$
becomes less than $\delta$, $k_{on}(r)$ in Eq. \ref{correctionkon} will vanish when $r<\delta$
and equal $k_{on}^{0}$ otherwise, as in the BR analysis. For a slowly varying repulsive $U(r)$,
force balance on the obstacle implies that typical values of $r$ will satisfy $dU/dr \simeq -F_{ext}$. Then
$U(r-\delta )-U(r) \simeq F_{ext}\delta$, so that $k_{on}$  is reduced by the familiar $\exp{(-F_{ext} \delta/k_BT)}$ factor.
However, in the case of an interaction potential with a deep narrow well, the results
can be quite different. For vanishing external force on the obstacle, basing the slowing on the
average obstacle force will give no correction. However, Eqs. \ref{correctionkon} and
\ref{correctionkoff} will give corrections to both $k_{on}$ and $k_{off}$.  The filament tip will generally be
near the bottom of the well. Therefore both $U(r-\delta )-U(r)$ and $U(r+\delta )-U(r)$ are positive,
so that $k_{on}$ and $k_{off}$ are reduced.

\subsection{Numerical results for different potentials and finite $D_{obst}$}

We calculated the force-velocity relations for a range of filament-obstacle interaction potentials described above,  
including  ``pusher" and ``single-well" potentials (Figure\,\ref{Potentials}a), and potentials having positive or negative Gaussian spikes as well as a double-well potential (Figure\,\ref{Potentials}b). The key parameter values are given in Table \ref{pars}. The free-filament polymerization rate $k_{on}^0$
is taken for a concentration of $1 \mu M$ actin with an on-rate constant of $11.6 \mu M^{-1} s^{-1}$ \citep{pollard1986rate}. 
Because the fractional error of the off-rate $k_{off}^0$ measured in Ref. \citep{pollard1986rate} is much larger than that of the on-rate constant,
we have assigned it a rough estimate of $1~s^{-1}$ corresponding to the general range of values in the literature. 
The obstacle diffusion coefficient $D_{obst}$ is taken as that of a sphere of radius $R = 5 \mu m$,  
using the Stokes relation $D_{obst} = k_BT/6 \pi \mu R$, where the viscosity $\mu$ is taken as that of cytoplasm, assumed to have a value $8.9 \times 10^{-3} Pa \cdot s$ ten times larger than that of water. This corresponds to relatively rapid
\begin{table}[h]
\caption{\label{table}Symbol definitions and parameter values.}
\begin{ruledtabular}
\begin{tabular}{clc}
Symbol & Definition & Value \\
\hline
$\delta$ & Actin step size & 2.7 nm \\
$D_{obst}$ & Obstacle diffusion coefficient& 5000 nm$^{2}$/s \\
$D_{tip}$ & Filament tip diffusion coefficient & $5\times 10^4$ nm$^{2}$/s \\
$\Delta t$ & Simulation timestep & $10^{-8}$s \\
$k_{on}^{0}$ & Free filament polymerization rate & 11.6 s$^{-1}$ \\
$k_{off}^{0}$ & Free filament depolymerization rate & 1 s$^{-1}$ \\
$F_{stall}$ & Filament stall force & 3.74 $pN$ \\
$U(r)$ & Potential of interaction between filament tip and obstacle & varies \\
$F(r)$ & Force exerted on obstacle by filament & varies \\
$F_{load}$ & External force on obstacle & varies \\
$v_{growth}$ & Filament growth velocity & varies \\
$r$ & Gap between filament tip and obstacle & varies \\
$z_{obst}$ & Obstacle z coordinate (height) & varies \\
\end{tabular}
\end{ruledtabular}
\label{pars}
\end{table}
diffusion according to the measure $D_{obst}/k_{on}^0\delta^2 \simeq 50$ \citep{peskin1993cellular}. 
Because this ratio is the key factor controlling the polymerization behavior, our results could also be taken
to describe, for example, a system with faster polymerization and faster diffusion. 
The effects of using lower values of 
$D_{obst}$ are described in Appendix B. The value of filament tip diffusion coefficent $D_{tip}$ (used in the calculations in Appendix A) 
is unknown. Because the moving part of the filament is much smaller than the sphere that we treat as an obstacle, 
we choose $D_{tip}$ to be $10$ times larger than $D_{obst}$, fast enough to ensure that filament-tip
fluctuations are much faster than obstacle fluctuations. We show results for a rigid filament growing at perpendicular incidence.
Results for a fluctuating filament tip, growing at oblique incidence, are given in Appendix A. 

Figure~\ref{TimeC} shows sample time courses of  filament length and obstacle position at zero load for two different force fields. The black curves correspond to a
``soft" repulsive potential (see Figure \ref{Potentials}a). The filament grows at roughly the free-filament rate and the obstacle has excursions of $100$ nm or more away
from the filament tip. The
blue curves correspond to a force field with an attractive well (Figure \ref{Potentials}a) of depth $5~k_BT$. Here the growth is slower
by about 50\%. The obstacle excursions are smaller on average. Although there are several ``mini-excursions" of tens of nm,
the obstacle returns to the filament. We consider this case to correspond to transient attachment of the
filament to the obstacle. However, later in the time course (at around $11$ sec), the obstacle has an excursion similar to that
seen for repulsive potential. Eventually, the filament will catch up to the obstacle and the excursions will diminish. 
\begin{figure}
\centering
\includegraphics[width=0.5\textwidth]{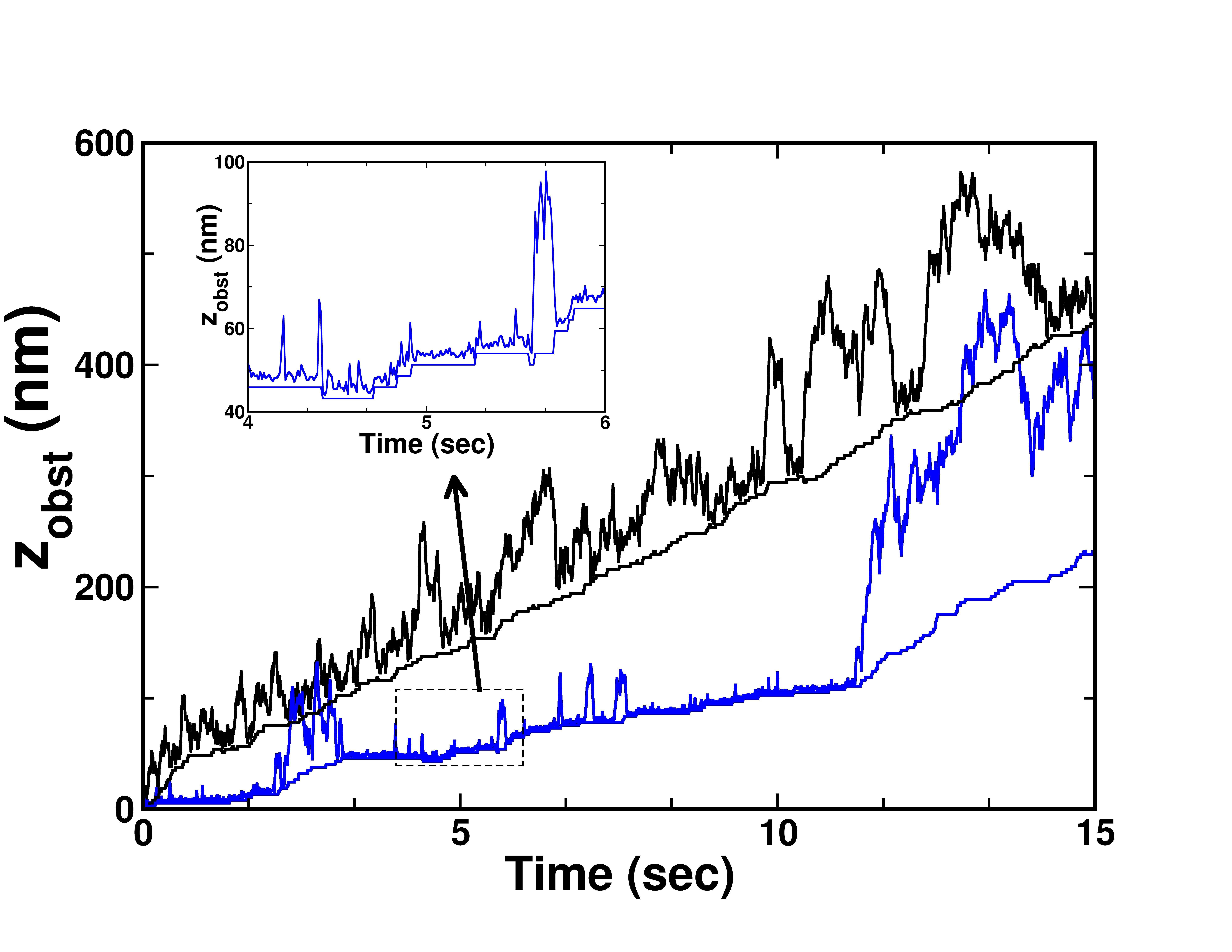}
\caption{Time course of filament length and obstacle motion for zero external force. Upper black curve is obstacle motion time course for a repulsive filament-obstacle interaction potential  (corresponding parameter values from Eq.~\ref{U1} are: A=54.7\,$pN$$\cdot nm$, B=0, and $\kappa_1$=0.9\,$nm^{-1}$) and the lower black curve is the filament height vs. time. Upper blue curve is the time course of the obstacle motion for an potential with an attractive well of depth $5~k_BT$ (parameter values from Eq.~\ref{U1} are: A=71.6\,$pN$$\cdot nm$, B=35.8\,$pN$$\cdot nm$, $\kappa_1$=0.9\,$nm^{-1}$, and $\kappa_2$=0.3\,$nm^{-1}$) with lower blue curve showing the filament height. Inset shows discrete polymerization and depolymerization steps, as well as the obstacle fluctuations against the filament tip.}
\label{TimeC}
\end{figure}

\begin{figure}
\centering
\subfigure[]{
\includegraphics[width=0.45\textwidth]{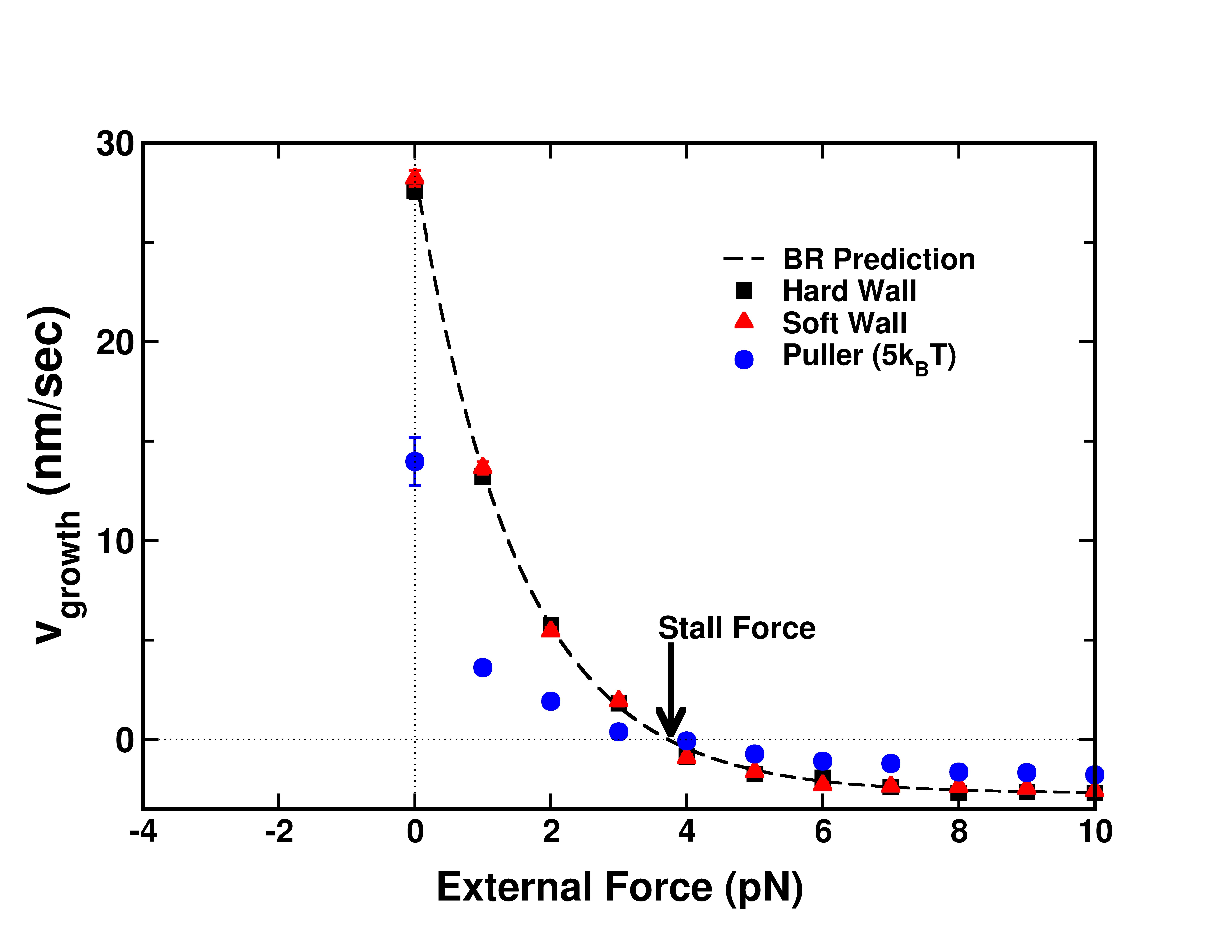}}
\qquad
\subfigure[]{
\includegraphics[width=0.45\textwidth]{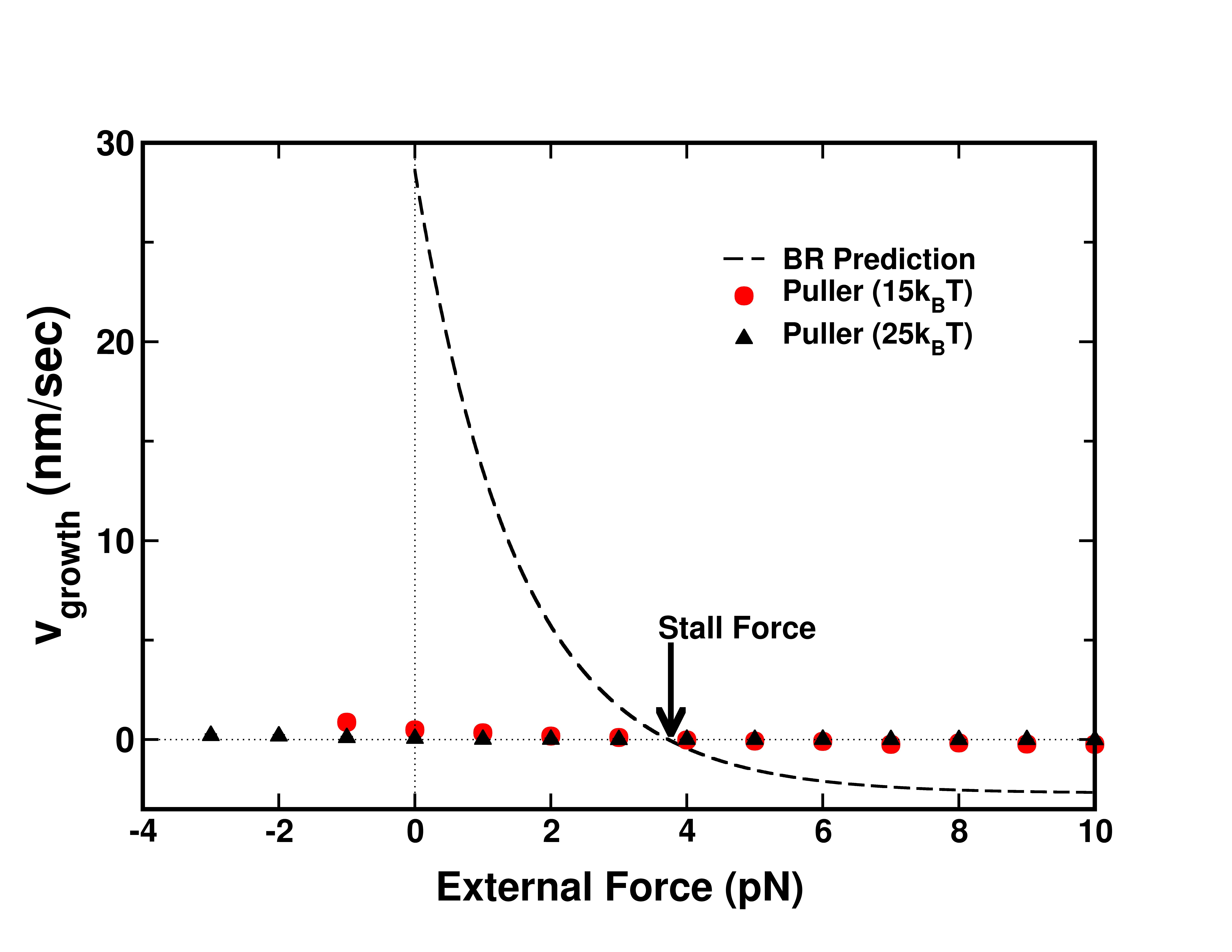}}
\qquad
\subfigure[]{
\includegraphics[width=0.45\textwidth]{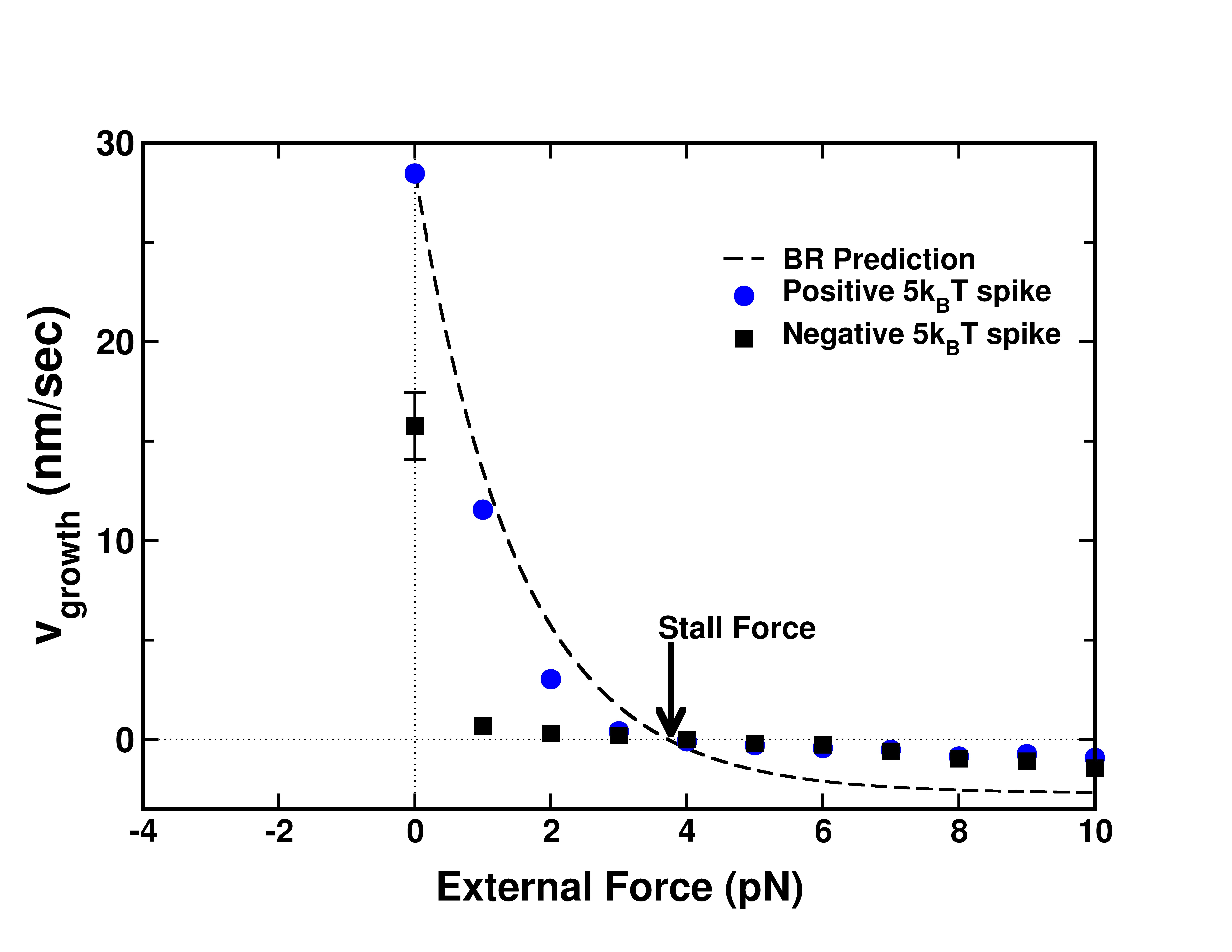}}
\qquad
\subfigure[]{
\includegraphics[width=0.45\textwidth]{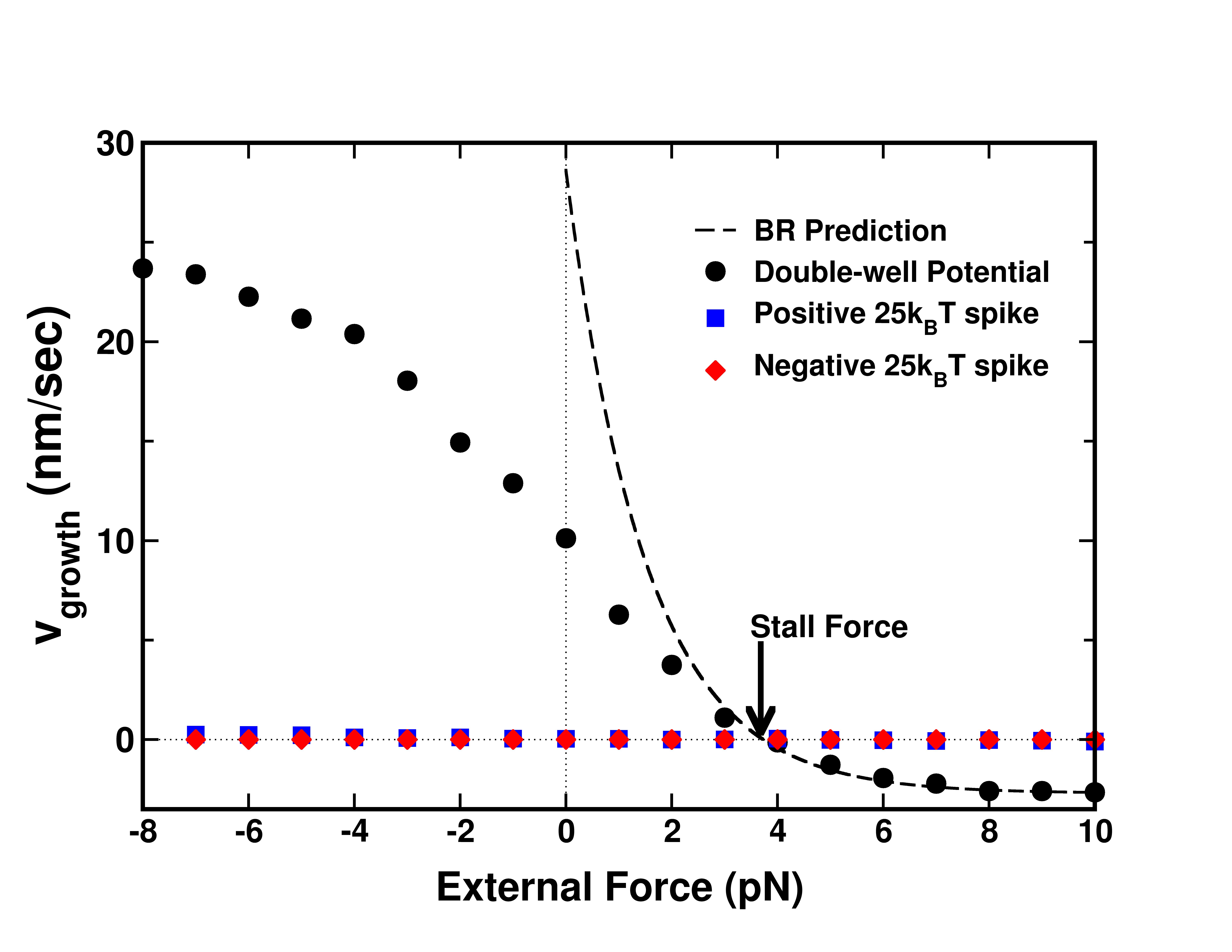}}
\caption{Simulated force-velocity relation for several different forms of filament-obstacle interaction potential. Error bars are smaller than the symbols, except indicated; and they are calculated by finding the standard deviation of the mean for growth velocity, from 20 different simulations of 20 seconds. The dashed line is the prediction of the BR model at large diffusion constant values for obstacle. Data points for pullers are shown out to forces where they detach from the obstacle. Frames (a) and (b) are for potentials in Figure \ref{Potentials}a, while (c) and (d) are for potentials in Figure \ref{Potentials}b.}
\label{F_V}
\end{figure}

Figure \ref{F_V} shows the calculated force-velocity relations for the range of potentials considered. 
For all the potentials, the growth velocity lies at or below the BR prediction. 
This is expected from the assumption (Eqs. \ref{correctionkon} and \ref{correctionkoff}) that $k_{off} \le k_{off}^0$:
Eq. \ref{hill} implies that ${\bar k}_{on} - {\bar k}_{off} = k_{on}^0 \exp{(-F\delta/k_BT)}f(F)- k_{off}^0 f(F)$, where $f(F) \le 1$. 
Then ${\bar k}_{on} - {\bar k}_{off}=f(F)[k_{on}^0 \exp{(-F\delta/k_BT)}- k_{off}^0]$, below the BR prediction.
For all purely repulsive potentials, the force-velocity relation is essentially indistinguishable from the BR prediction. 
This again follows from Eq. \ref{hill}, \ref{correctionkon}, and \ref{correctionkoff}. For repulsive potentials 
Eqs. \ref{correctionkon} and \ref{correctionkoff} imply that $k_{off}(r)= {\bar k}_{off} = k_{off}(0)$, so that
${\bar k}_{on}= k_{on}^0 \exp{(-F\delta/k_BT)}$ and the BR relation holds.

Simple-well potentials, as well as potentials having negative spikes and double wells (Figure\,\ref{Potentials}b),  lead to slowing of growth relative to the BR model in the positive (pushing) force regime. 
For a well depth of $5~k_BT$, the zero-force velocity is about half of the free-filament velocity, larger than might have been expected from the 
$\exp{\{ -[U(r-\delta )-U(r)]/k_BT \}}$ factor in Eq. \ref{correctionkon}, which is approximately $\exp{(-5})$ in the case of the Gaussian spike when the
obstacle is near the bottom of the potential well. The reason for the faster growth is that in the absence of external force, the obstacle
is outside the well a substantial fraction of the time. 

To see this, we assume fast obstacle diffusion and exploit the fact that the growth of the filament tip at velocity $v$ toward
the obstacle has the same effect on the distance distribution $P(r)$ as a weak external effective force $F_{eff} = k_BT v/D_{obst}$
\citep{peskin1993cellular}. This holds because the equation of motion for $P(r)$ in the
presence of filament growth (taken to have a constant velocity, and using the Einstein relation) is 
$\partial P/\partial t = D_{obst} \partial^2 P/\partial r^2+ [F(r)D_{obst}/k_BT+v]\partial P/\partial r$, while the equation of
motion in the presence of a constant force $F_{eff}$ is $\partial P/\partial t = D_{obst} \partial^2 P/\partial r^2+ \{[F(r)+F_{eff}]D_{obst}/k_BT\}\partial P/\partial r$.

We assume the obstacle to be in either the region of the potential well, where we ignore polymerization,
or in the region outside the well, where it polymerizes at the free-filament velocity $v_0$. 
This picture is most applicable to narrow wells, such as the ``Negative Spike" treated in Figure \ref{F_V}c.		
The polymerization rate is then 
$v/v_0 = Z_{free}/(Z_{well}+Z_{free})$, where $Z_{well}$ is the contribution to the obstacle's partition function from the well region and 
$Z_{free}$ is the contribution from outside the well. We take
\begin{equation}
Z_{free} = \int_0^\infty exp{(-F_{eff} \cdot r/k_BT)}dr=\frac{k_BT}{F_{eff}} = \frac{D_{obst}}{v}   
\end{equation}
where we have taken the integral to extend from $0$ to $\infty$ for mathematical simplicity, which is valid as long as $F_{eff}$ is weak. 
This gives
\begin{equation}
\frac{v}{v_0} = \frac{1}{1+Z_{well}/Z_{free}} = \frac{1}{1+Z_{well}v/D_{obst}}
\label{zwelleq}
\end{equation}
The solution to this equation is
\begin{equation}
\frac{v}{v_0} = \frac{2}{1+\sqrt{1+4\eta}},
\label{vfreeanal}
\end{equation}
where $\eta = v_0 Z_{well}/ D_{obst}$. 
Treating the ``Negative $5 k_BT$ Spike" as a square well of depth $5 k_BT$ and width $2$ nm, we obtain 
$Z_{well} = 2~nm\cdot e^5$ and $v/v_0= 0.52$, roughly consistent
with the numerical results in Figure \ref{F_V}c. 
The growth velocity given by Eq. \ref{vfreeanal}
is appreciable only if $\eta \lesssim 1$, so that $D_{obst} \gtrsim v_0 Z_{well}$.
Therefore, even though $D_{obst}/k_{on}^0\delta^2 >> 1$, the growth velocity depends strongly
on the diffusion coefficient.

For the potentials with $5 k_BT$ wells, the velocity also decays more rapidly with opposing force than the BR relation predicts. 
For example, the ``5 k$_B$T" curve in Figure \ref{F_V}a drops by
nearly 75\% already at 1 pN force, while the BR model and repulsive force fields drop by only about 50\%. For the ``Negative 5 k$_B$T" spike in Figure \ref{F_V}c,
the drop is practically down to zero.  This rapid drop occurs because the external force reduces 
the statistical weight of the free region, which becomes
$Z_{free}= k_BT/(F_{eff}+F_{ext})$. 
Substituting this into Eq. \ref{zwelleq} gives
\begin{equation}
\frac{v}{v_0} = \frac{2}{1+{\tilde f}+\sqrt{(1+\tilde f)^2+4\eta}},
\label{fvanal}
\end{equation}
where $\tilde f=Z_{well} F_{ext}/k_BT$. Even
small values of $F_{ext}$ can affect the velocity strongly, because of the $Z_{well}$ factor in $\tilde f$.  
Solving Eq. \ref{fvanal}, again using a square well of depth $5 k_BT$ and width $2$ nm, we find 
$v/v_0= 0.014$ at $F_{ext}= 1$ pN, consistent with the very rapid drop seen in Figure \ref{F_V}c. 
When diffusion is very rapid, $\eta \rightarrow 0$, and 
\begin{equation}
v/v_0 \simeq 1/(1+\tilde f ). 
\label{vftilde}
\end{equation}
Taking $(1/v)(-dv/d{F_{ext}}) \ge 2\delta/k_BT $ at zero force (twice the BR value) as a definition of rapid decay with force, we find that
rapid decay will occur when $Z_{well} \gtrsim 6$ nm. 

For the double-well potential (Figure \ref{F_V}d), the velocity
decays less rapidly relative to its $F=0$ value than the BR prediction (although the magnitude of $v_{growth}$ is always smaller than the BR result); in fact the drop is almost linear. 

\begin{figure}
\centering
\subfigure[]{
\includegraphics[width=0.45\textwidth]{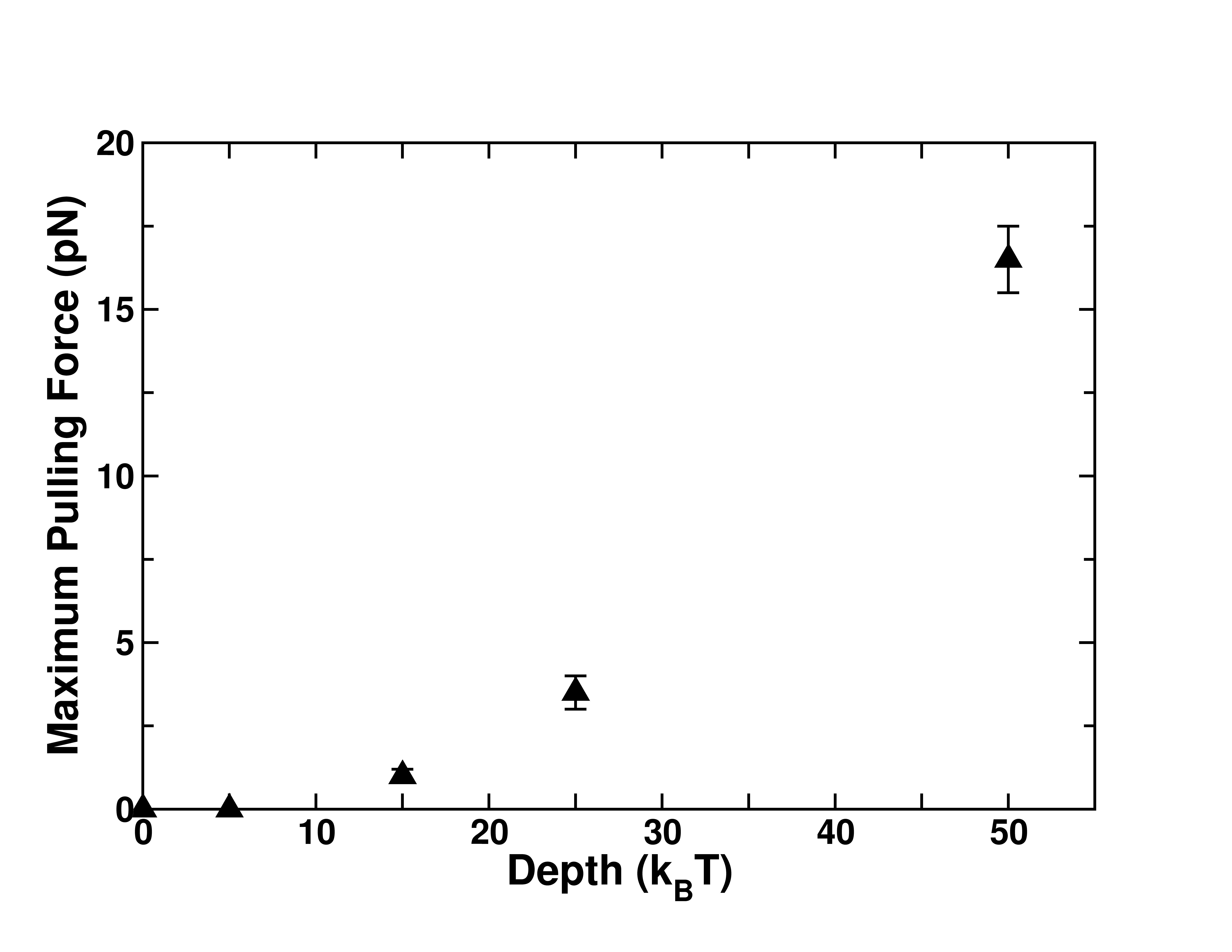}}
\qquad
\subfigure[]{
\includegraphics[width=0.45\textwidth]{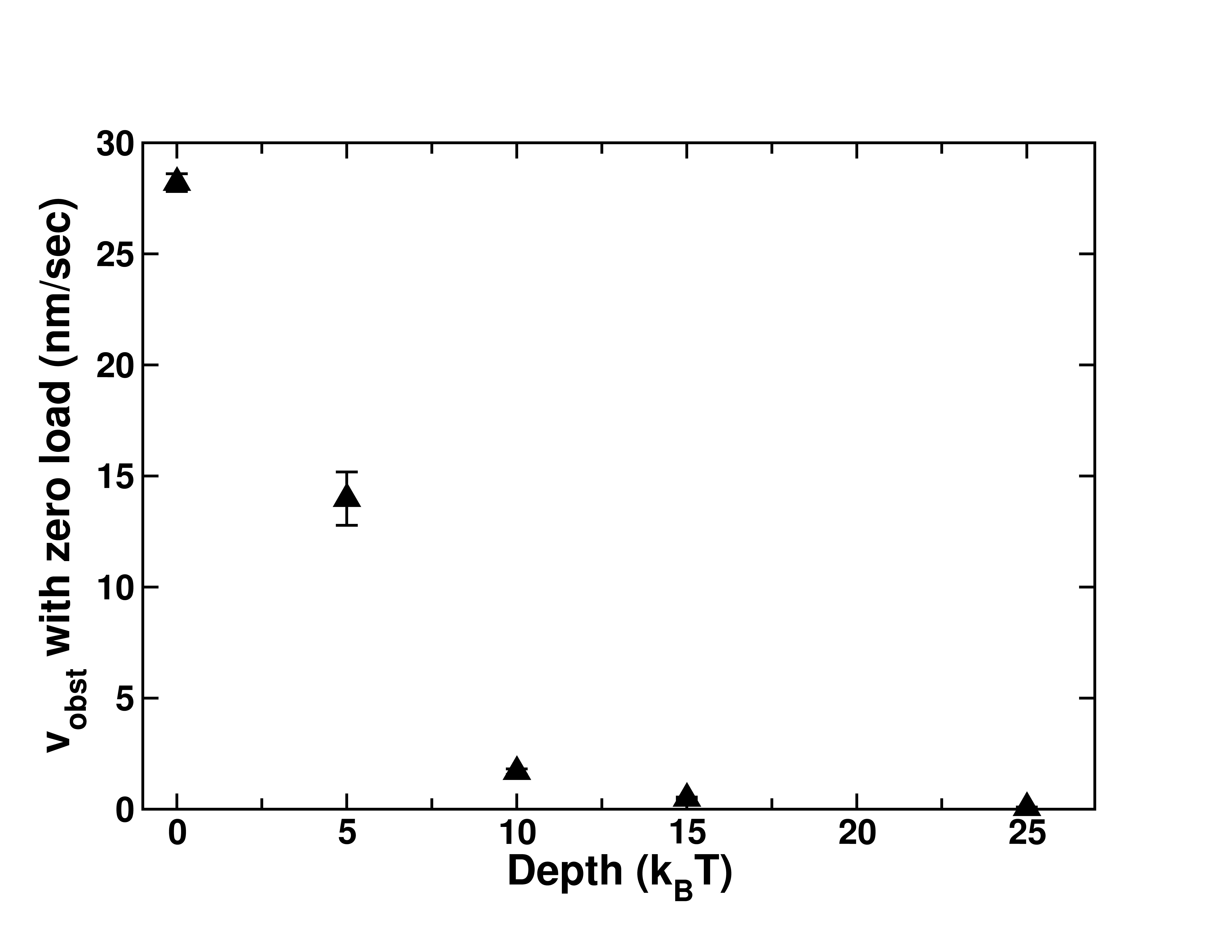}}
\caption{Effect of binding strength of single-well potential on a) the maximum pulling force allowing attachment up to 20 seconds, and b) the growth velocity at zero external load.}
\label{results}
\end{figure}

Another feature of the systems with potential wells is that they can polymerize processively under pulling (negative) force, over a limited time \citep{zhu2006growth}. 
As expected physically, there is a tradeoff between maximum sustainable force and polymerization rate. High pulling force enhances
polymerization, but at the same time accelerates detachment of the obstacle from the filament. 
For a 5$k_BT$ potential depth, almost no pulling force can be sustained over 20 seconds (Figure \ref{F_V}a). But for a 15$k_BT$ depth  (Figure \ref{F_V}b), a force of about 1 pN can be sustained over 20 seconds. The growth velocity in this case is about 80\% greater than the zero-force value, but  much lower than for the $5~k_BT$ case. 
For the 25$k_BT$ potential still larger pulling forces can be sustained, at an even smaller growth rate. Figs.\,\ref{F_V}(c) and Figure \ref{F_V}d show 
that the same trade-off occurs for potentials with spikes. 
The trade-off is explicitly illustrated in Figure \ref{results}. Frame a) shows how the maximum pulling force that allows a 20-second attachment period, depends on the depth of the potential well. Frame b) shows how the growth velocity depends on the well depth. These two plots show that large sustainable pulling forces on the obstacle (greater than about 1 pN) come at the expense of greatly reduced growth velocity. 
Pullers with a well depth of about 5$k_BT$ grows only half as fast as a free filament, but can sustain essentially no pulling force.  If the depth is larger than about 15$k_BT$, which is needed to sustain pN forces, growth almost completely stops. 
The only form of potential used here that achieves a reasonable growth rate at a substantial pulling force is the double-well potential (Figure \ref{Potentials}b).
As shown in Figure \ref{F_V}d, this potential
eliminates the trade-off between sustainable force and polymerization rate that is seen with other potentials. This is because the well is wide enough that polymerization can occur inside even a very deep well. If the obstacle is pulled toward the large-$r$ end of the well, 
$U(r-\delta)$ and $U(r)$ do not differ greatly, so there is no significant slowing in Eq. \ref{correctionkon}. 

\section{Discussion}

Our main findings are the following:

\begin{itemize}
\item{The ratio of the on- and off-rates depends on the details of the interaction potential
with the obstacle on a nanometer scale (see Eq. \ref{rigorous}). This implies that even at zero force, a deep well in the interaction potential
can dramatically slow both polymerization and depolymerization. Polymerization is slowed because of the high-energy configuration
assumed by the filament tip after addition of a new subunit, while depolymerization is slowed because removal of an existing subunit
also leaves the filament in a high-energy configuration. Eq. \ref{rigorous} is automatically satisfied
by calculations implementing the BR approximation that $k_{on}(r)=0$ when
the filament tip is within a distance $\delta$ of a hard obstacle with no attractive well, but $k_{on}(r)=k_{on}^0$ 
beyond that distance. 

Not all single-filament calculations in the literature treating obstacle motion explicitly have satisfied
the constraint of Eq. \ref{rigorous}. We note that Eq. \ref{rigorous} does not apply to
calculations such as as those of Refs. \citep{carlsson2000force}, which ignored depolymerization.
Ref. \citep{burroughs2006growth} assumed a variation of the on-rate resulting from the interaction 
potential with the obstacle, but it did not appear to satisfy Eq. \ref{rigorous}.  On the other hand, 
Ref. \citep{zhu2006growth} used a relationship very similar to Eq. \ref{rigorous}, in which 
the exponential factor corresponded to the energy difference between binding in two different positions in a double-well
potential. 

Some works treating force generation by actin networks stochastically have also treated smoothly varying
tip-obstacle interactions \citep{carlsson2001growth,schaus2007self,schaus2008performance,hu2010mechano,zhu2012mesoscopic,hu2013molecular,mund2018systematic}. 
In all of these works, 
actin polymerization was slowed by the obstacle. But none of the treatments satisfies Eq. \ref{rigorous}
exactly. Ref. \citep{carlsson2001growth} used an approximate version in which the $U(r)$ term in
the exponential of Eq. \ref{rigorous} is ignored.  In Refs. \citep{schaus2007self} and \citep{schaus2008performance},
the mechanical energy, including the membrane deformation/position and states of actin filament bending,
was minimized after each polymerization event. To evaluate the on-rate, the
energy was reminimized with a hypothetical next subunit in place. The resulting energy difference
$\Delta U$ was used in a Boltzmann factor slowing the polymerization. This approach is similar in spirit 
to Eq. \ref{rigorous}, but differs in that the energy is minimized in calculating $\Delta U$. In the context of
the systems treated here, this would correspond to placing the obstacle, before and after the addition of
the new subunit, at the minimum of the potential $U(r)+F\cdot r$. Then $\Delta U = F \delta$,  giving a 
slowing of the growth velocity of a single filament by a factor of $\exp{(-F \delta/k_B T)}$. 
As found above, this is correct provided there is no attractive well in the potential.  

Refs. \citep{hu2010mechano,zhu2012mesoscopic,hu2013molecular,mund2018systematic} used criteria based on
force rather than energy difference. In such approaches,
even plausible assumptions regarding the slowing of actin polymerization can lead to 
substantial errors in the stall force. For example, one
can assume \citep{mund2018systematic} that a filament tip experiences a linear force 
$F(r) = - k_c r \theta (-r)$ when in contact with the obstacle, and that
this force slows polymerization according to 
\begin{equation}
k_{on} (r) = k_{on}^0 \exp{[-F(r) \delta/k_BT]},
\label{riesslow}
\end{equation}
while $k_{off}(r)= k_{off}^0$ is not affected.  This corresponds to applying the thermodynamic result
Eq. \ref{hill}, for the averaged rate ${\bar k}_{on}$, to the instantaneous rate
$k_{on} (r)$. The force-velocity relation can be obtained exactly for this model in the
limit $D_{obst} \rightarrow \infty$. Using the same analysis as used to derive Eq. \ref{konbar},
but incorporating Eq. \ref{riesslow}, we obtain
\begin{eqnarray} 
{\bar k}_{on} & = &(k_{on}^0/Z)  \int_{-\infty}^{\infty}\exp{[-F(r) \delta/k_BT]}\exp{\{-[U(r)+F_{ext} \cdot r]/k_BT\}} dr \nonumber \\
&=& (k_{on}^0/Z)\left[ \sqrt{\frac{\pi k_BT}{2k_c}}\exp{\{[F_{ext}-k_c \delta]^2/2k_ck_BT\}}\{ 
1+{\rm erf} [(F_{ext}-k_c \delta )/\sqrt{(2k_ck_BT)}] \} + k_BT/F_{ext} \right] 
\label{konbarries}
\end{eqnarray}
where $U(r) = k_c \theta (-r)r^2/2$ and
\begin{equation}
Z =  \sqrt{\frac{\pi k_BT}{2k_c}}\exp{(F_{ext}^2/2k_ck_BT)}
\{ 1+{\rm erf} (F_{ext}/\sqrt{(2k_ck_BT}) \} + k_BT/F_{ext}.
\label{zries}
\end{equation}
The stall force for this model can be obtained by numerically locating the zero of ${\bar k}_{on} - {\bar k}_{off}$. 

In the limit of large stall force ($F_{ext} - k_c \delta >> \sqrt{2 k_c k_B T}$),  Eq. \ref{konbarries}
simplifies. The error functions in Eqs. \ref{konbarries} and \ref{zries} approach unity, and the $1/F_{ext}$ terms can
be ignored. Thus ${\bar k}_{on} \simeq k_{on}^0 \exp{(-F_{ext}\delta/k_BT)}\exp{(k_c \delta^2/2k_BT)}$. Solving for
${\bar k}_{on}= {\bar k}_{off}$ we obtain 
\begin{eqnarray}
F_{stall}&= &(k_B T/\delta)\ln{(k_{on}^0/k_{off}^0)}+ k_c \delta/2,
\label{fstallries}
\end{eqnarray}
which exceeds the thermodynamic stall force by the amount $k_c\delta/2$.
Numerically, both from Eq. \ref{konbarries} and from our stochastic simulations with a large but finite $D_{obst}$, 
we find comparable overestimates of the stall force in other parameter ranges. The overestimates are equally large in parameter ranges where
filament-tip fluctuations dominate. For the spring constant assumed in Ref.
\citep{mund2018systematic} we obtain a stall force of $50$ pN, about five times too high given
the assumed actin concentration. 
We emphasize that these results hold only when thermal motion of the obstacle or filament
tip is explicitly treated rather than being averaged out. 

Why does Eq. \ref{konbarries}, which appears to correctly implement the force dependence of actin polymerization by
equating the instantaneous rates to the known results (Eq. \ref{hill}) for the average rates, 
fail to obtain the correct stall force? The reason is that although the force $F(t)$ acting
on the tip of a filament polymerizing under
an average force $F_{ext}$ satisfies ${\bar  F}  =F_{ext}$, the fluctuating motion of the obstacle or 
the filament tip causes the tip to experience a range of forces
described by a distribution $p(F)$, so that 
\begin{eqnarray}
{\bar k}_{on} &=& k_{on}^0\int_{-\infty}^{\infty} p(F) \exp{(-F \delta/k_BT)} dF
\label{probf}
\end{eqnarray}
Since the exponential is convex upwards (the second derivative of $\exp{(-F \delta/k_BT)}$ is positive), ${\bar k}_{on}$ will exceed 
$k_{on}({\bar  F})$, leading to an overestimate of the stall force. If one assumes that 
$p(F) \propto \exp{[(F- {\bar F})^2/2\Delta F^2]}$ where $\Delta F^2$ is the variance of $F$, Eq. \ref{probf} shows that
${\bar k}_{on}= k_{on}^0\exp{[-{\bar F}\delta/k_BT]}\exp{[\delta^2 \Delta F^2/2(k_BT)^2]}$. Setting
this equal to $k_{off}^0$, one finds 
\begin{eqnarray}
F_{stall} &=& (k_B T/\delta)\ln{(k_{on}^0/k_{off}^0)}+ \Delta F^2 \delta/2k_BT. 
\label{fstallflucts}
\end{eqnarray}
For force fluctuations induced by thermal motion in a harmonic potential, $\Delta F^2 = \Delta r^2 k_c^2 =k_c k_B T$, and Eq. \ref{fstallflucts} 
is identical to Eq. \ref{fstallries}. On the other hand, if thermal motions of the filament tip or obstacle come from non-thermal 
sources, as in Refs. \citep{hu2010mechano,zhu2012mesoscopic,hu2013molecular}, 
force fluctuations could instead result from the geometrical constraints imposed on the individual filaments by the network structure. 
Then one might expect the bending-induced fluctuation of a given filament tip position to be $\lesssim \delta$; larger deformations would likely be evened out
by the differences in growth velocity. As described in Appendix A, one can estimate the spring constant
of a filament in the direction of motion as  $k_{bend} \simeq 0.5~\frac{pN}{nm}$.
Then $\Delta F \sim k_{bend} \delta  \lesssim 1.35$ pN, and the overestimate in Eq. \ref{fstallflucts} is
$\lesssim 0.6$ pN, much smaller than for thermal force fluctuations.
} 
\item{Provided that $k_{off} (r) \le k_{off}^0$ and $D_{obst}$ is large, no type of filament-obstacle interaction potential 
leads to polymerization faster than a hard wall;
for monotonically decaying force fields the force-velocity relation is very near that 
for a hard wall, which is described well by the BR prediction. 
These results follow from Eq. \ref{hill} and are confirmed by the simulations. 
For intermediate values of $D_{obst} \sim k_{on}^0 \delta^2$, we find that softer potentials accelerate growth
slightly (see Figure \ref{SlowDiffusion}).

But when $D_{obst}$ is very small, a soft potential can accelerate polymerization substantially.
This was found in the simulations of
\citep{burroughs2006growth}, where a soft obstacle accelerated polymerization by 
about 100\% using $D_{obst} = 0.0016 k_{on}^0 \delta^2$.
To understand this physically, consider the case of zero external force. 
In the limit of a slowly varying potential, it is legitimate to ignore
the randomness in polymerization and treat filament growth as occurring at a constant velocity. As 
described in Section 3.2, obstacle drag can then be included
via an effective force $F_{eff}=k_B T v_{obst}/D_{obst}$ \citep{peskin1993cellular}.
After a sufficiently long time, the probability distribution $P(r)$ will settle into a steady-state form
$P(r) = \exp{\{ -[U(r)+F_{eff} \cdot r]/k_BT}\}/Z$, where $Z$ is the corresponding partition function.
Then following the derivation of Eq. \ref{konbar}, and taking $f(r)=1$ in Eq. \ref{kform} so that ${\bar k}_{off} = k_{off}^0$,
\begin{eqnarray}
{\bar k}_{on} & = & (k_{on}^0/Z) \exp{( -F_{eff}\delta/k_BT )} \int_{-\infty}^{\infty} \exp{[ -U(r)/k_BT ]} \exp{[ -F_{eff} \cdot r/k_BT ]}dr \nonumber \\
& = &  \exp{( -F_{eff}\delta/k_BT )}  k_{on}^0
\label{konbarsoft}
\end{eqnarray}
Since $v_{obst} = ({\bar k}_{on}- k_{off}^0)\delta$, it follows that 
$v_{obst} = (D_{obst}x/\delta - k_{off}^0 \delta)$, where $x$ satisfies the transcendental equation
$~x /{\tilde k}_{on}  =\exp{(-x)} \exp{({\tilde k}_{off})}~$ and we define ${\tilde k_{on,off}} = (k_{on,off}^0\delta^2/D_{obst})$.  
Therefore 
\begin{eqnarray}
v_{obst} = &(D_{obst}/\delta)\{W[{\tilde k}_{on}\exp{({\tilde k}_{off})}] - {\tilde k}_{off}\} &  \label{vanal1}\\
 \simeq  & (D_{obst}/\delta)\ln{({\tilde k}_{on}/{\tilde k}_{off})} & ({\tilde k}_{on} > > 1, ~ {\tilde k}_{off} > > \ln{( {\tilde k}_{on})})  \label{vanal3}\\
 \simeq  & (D_{obst}/\delta) [ \ln{({\tilde k}_{on})} - \ln{(\ln{\tilde k}_{on})} ] & ({\tilde k}_{on} > > 1,~ {\tilde k}_{off} \rightarrow 0)
\label{vanalslow}
\end{eqnarray}
where the Lambert function $W$ is the inverse of the function $x\exp{(x)}$. We 
have used the asymptotic expansion $W(x) \simeq \ln{(x})- \ln{[\ln{(x)} ]}$, valid when 
$x$ is large (very slow diffusion). 

In cases of very slow obstacle diffusion, the simplified form Eq. \ref{vanal3} will hold
if the ratio between the on- and off-rates is moderate. 
The physical content of this result is clarified by considering the work done
in moving the obstacle. The work done per added subunit
is $(v_{obst}k_BT/D_{obst})\delta$ where the first term is the force according to
the Einstein relation. From Eq. \ref{vanal3} this equals $k_BT \ln{(k_{on}^0/k_{off}^0)}$,
which is the free energy released per added subunit \citep{hill2012linear}. Thus 
the process is 100\% efficient in that all of the
free energy of polymerization is used to push the obstacle against the drag force.
If $(k_{on}^0/k_{off}^0)$ is large, 
the velocity in Eq. \ref{vanal3} could exceed the hard-wall limit $2D_{obst}/\delta$ 
\citep{peskin1993cellular} substantially. 
For example, for a free-actin concentration of $10 \mu M$, our parameters would predict that
$(k_{on}^0/k_{off}^0)\simeq 100$, and the logarithm is greater than $4$. Then the hard-wall 
limit would be exceeded by more than a factor of 2.  

The simulations of Ref. \citep{burroughs2006growth} treated the case ${\tilde k}_{on} = 6.67 \times 10^2$
and ${\tilde k}_{off}=0$.  In this case, Eq. \ref{vanal1} gives $v_{obst}=4.9 D_{obst}/\delta$,
while Eq. \ref{vanalslow} gives $v_{obst}=4.6 D_{obst}/\delta$, both about a factor of two
above the hard-wall limit and comparable with the value of $4.4 D_{obst}/\delta$
found in Ref. \citep{burroughs2006growth}. This suggests that the analytic theory captures 
the key effects in these simulations.  
}
\item{For relatively shallow attractive potentials the velocity can be a substantial fraction of
the free-filament velocity at zero force, but decay rapidly with opposing force. This finding may
help explain the results of experiments \cite{footer2007direct} studying small number of actin filaments 
growing against a hard wall, if the filament tips are weakly bound to the wall.  
The filaments propelled acrosomes attached to beads held in an optical trap (backwards). 
The filaments/acrosome/bead are moving rather than the obstacle, so as discussed above it is
their motion that is considered.  The growth velocity was found to drop off much
more rapidly than expected from the BR model at forces of a few tenths of a $pN$ per filament, 
especially at a $2 \mu M$ actin concentration. 
We cannot treat their many-filament system within our model, so we consider 
a single filament growing against an obstacle with a force of a few tenths of a $pN$. 
Because the bead is trapped in potential well of spring constant $k_c \simeq 0.008 pN/nm$,
the ``free" contribution $Z_{free}$ is different from that calculated in Sec. 3.2. 
The energy of the bead in the potential well, displaced a distance $r$ from the minimum, is 
$k_c r^2/2$. 
Thus $Z_{free} = \int_{0}^{\infty}\exp{(-F_{ext} \cdot r)} \exp{(-k_c r^2/2)} dr$. 
Provided that $F_{ext} \ge \sqrt{k_c k_BT} = 0.2 pN$, the reduction in $Z_{free}$ from 
$F_{ext}$ will exceed that from $k_{c}$, and it is reasonable to take $Z_{free} = k_BT/F_{ext}$ as in Sec. 3.2.
Then Eq. \ref{vftilde} applies, and a force of few tenths of a pN per filament could reduce 
the velocity by a large factor if $Z_{well} \ge 15 nm$. The validity of
Eq. \ref{vftilde} requires that $\eta = v_0 Z_{well}/D_{obst} << 1$. In Ref.
\citep{footer2007direct}, $v_0$ was less than $20 nm/s$, so a diffusion coefficient
$\ge 600 nm^2/s$ would be adequate.  


Experiments on whole cells  \cite{prass2006direct,zimmermann2012actin} have
shown that the growth velocity of lamellipodia drops very rapidly with opposing force. The velocity
decay found here may contribute to this effect, but only if the filaments are sufficiently long
to allow thermal fluctuations greater than $Z_{free} = k_BT/F_{ext}$ away from the obstacle. In this system, 
mechanical factors may be the dominant effect \citep{zimmermann2012actin}.}

\item{Sustaining strong pulling forces at a significant rate of polymerization requires a deep, broad well in the filament-obstacle interaction.
Actin filaments polymerizing under pulling force may have several functions. 
They could act as parts of force sensors, or as mechanical absorbers for rapidly generated forces from
myosin motors \citep{yu2017mdia1}. It is also believed that actin filaments in the central region of endocytic sites
in budding yeast exert pulling forces on the membrane \cite{wang2016actin,mund2018systematic}, and it is important to
know if filaments can polymerize sufficiently quickly to generate a gel in the pulling region that can sustain the large stresses
generated by the process.
The behavior of actin filaments under pulling forces has been addressed by thermodynamic arguments \citep{kozlov2004processive}
as well as kinetic models \citep{yu2017mdia1,kubota2017biphasic} and simulations \citep{bryant2017computational,zhu2006growth}. 
Most of the calculations have predicted acceleration of polymerization by pulling force, but Ref. \citep{bryant2017computational}
found a competition between different conformational factors that could either slow or speed polymerization. 
Experimental studies \citep{courtemanche2013tension,kubota2017biphasic,yu2017mdia1,cao2018modulation} have shown that polymerization under
pulling forces of several pN is possible if actin filaments are linked to the obstacle via formins. They suggest, on the whole,
that pulling force accelerates polymerization if rotational constraints are absent. These models have included effects not explicitly included here,
such as conformational changes of formins at the actin filament tip. Our finding that for narrow wells
polymerization is incompatible with the ability to sustain large forces implies that in the
systems where this phenomenon occurs, the interaction between the actin filament and the obstacle must have
a broad minimum. Our models are too simple to quantitatively describe the three-dimensional geometry of a formin-tipped
actin filament, but the broad minimum in our ``double-well" potential may approximately mimic the conformational flexibility that 
appears to be at the heart of the phenomenon. 
}
\end{itemize}

The calculations described here make several major approximations, including the treatment of just a single filament, and the modeling
of the filament-obstacle interaction via simple potential energy functions. To make direct contact with experiments will require
more complex calculations for many-filament systems. The present results can help make progress toward this goal 
by informing multiscale calculations such as those of Ref. \citep{mogilner2002regulation, mogilner2005physics,rubinstein2005multiscale, maree2006polarization, 
ryan2017cell,ditlev2009open, joanny2009active,kim2009continuum,craig2012membrane,zimmermann2012actin,adler2013closing,khamviwath2013continuum,carlsson2014force,barnhart2017adhesion,camley2017crawling,mueller2017load}, which treat force generation by multifilament systems
using a variety of approximations to include the single-filament force-velocity relation. The present results will
provide useful guidance, especially in cases where different types of filament-obstacle interactions are present in the same system.

\begin{acknowledgments}
This work was supported by the National Institute of General Medical Sciences (https://www. nigms.nih.gov) under Grant R01 GM107667 and the National Science Foundation (https:// www.nsf.gov) under Grant Agreement CMMI:15-458571. 
\end{acknowledgments}


\newpage

\centerline{\bf \large APPENDIX}

\setcounter{page}{1}

\bigskip

{\bf The effect of the filament-obstacle interaction on the force-velocity relation of a growing biopolymer}

\bigskip

\centerline{F.~Motahari and A.~E.~Carlsson}
\bigskip
\bigskip
\bigskip
Here we extend the results beyond the simplifying approximations made in the body of the paper, by including
filament-tip fluctuations, oblique incidence, and slower diffusion.

\appendix

\renewcommand\thefigure{\thesection.\arabic{figure}}    

\section{Filament-Tip Fluctuations and Oblique Incidence}
\setcounter{figure}{0}    

\begin{figure}[H]
\centering
\includegraphics*[width=0.5\textwidth]{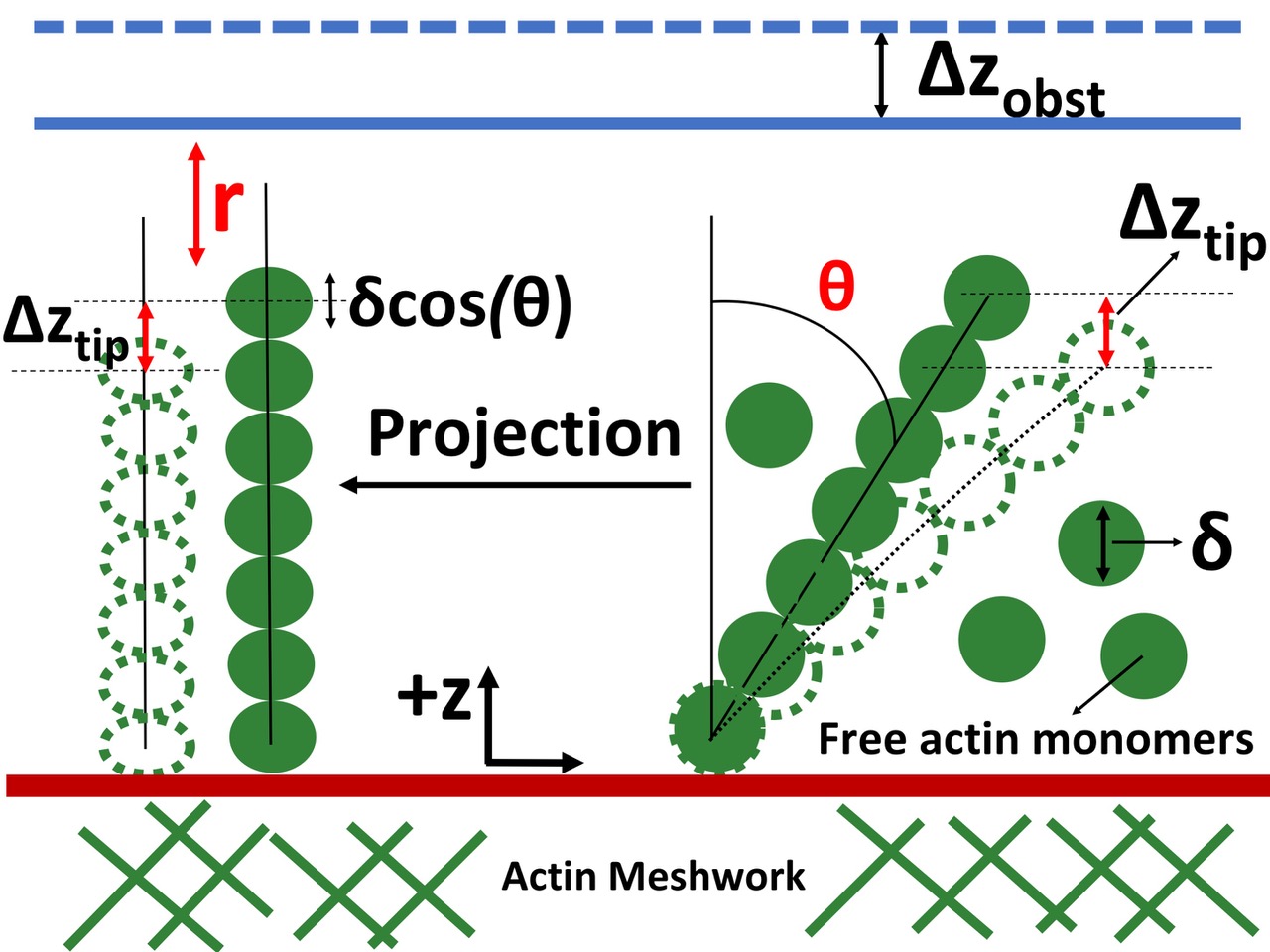}
\caption{Schematic of model of filament at oblique orientation with incident angle $\theta$. $\Delta z_{tip}$ the filament tip fluctuation.}
\label{model_ob}
\end{figure}

We treat oblique incidence together with filament-tip fluctuations, since these
fluctuations are much greater at oblique incidence than at
perpendicular incidence.  We use an incidence angle of $\theta = 45^o$ as in Figure \ref{model_ob}.
The filament-tip fluctuations are modeled by a variable $z_{tip}$ describing the 
deflection of the tip, assumed to move according to Brownian dynamics in a quadratic potential well:
\begin{equation} \label{z_memb_fluctuation}
\Delta z_{tip} = \alpha\ensuremath{'}\sqrt{24\Delta t}\sqrt{D_{tip}} + \frac{D_{tip}}{k_BT}\Delta t [ -F(r) - k_{bend}\cdot z_{tip} ]
\end{equation}
Here $D_{tip} = 5\times10^4\,\frac{nm^2}{sec} = 10D_{obst}$ is the filament tip diffusion coefficient. The true value of
$D_{tip}$ is probably greater than this, since the part of the filament free to bend is much smaller than
the $5 \mu m$ obstacle that we consider. However, using the actual value would render the simulations extremely
demanding. For this reason we have chosen a value an order of magnitude larger than $D_{obst}$, so that
the tip fluctuations will equilibrate on time scales much faster than that of obstacle motion.
The variation of the deflection $z_{tip}$ is limited: 
the tip can not bend down past its own base or up so far that its height from the base exceeds the filament length.
The time step $\Delta t = 10^{-9} sec$ is is chosen so that filament tips will move much less than the subunit
size in one time step. The tip bending stiffness  $k_{bend}$ is
obtained \citep{lautrup2005physics} as $k_{bend}= 3 k_BT L_p/L^3\sin^2{\theta} = 0.5~\frac{pN}{nm}$, 
where $L_p \simeq 20 \mu m$ \cite{gittes1993flexural,isambert1995flexibility} is the persistence
length  and $L$ is the filament length, which we take to have a typical lamellipodium value of $100$ nm. 
Finally $\alpha\ensuremath{'}$ is a random number uniformly distributed between $-\frac{1}{2}$ and $\frac{1}{2}$, 
so that $<\alpha\ensuremath{'}\,^2> = \frac{1}{12}$. 
Consecutive time steps are uncorrelated.  

\begin{figure}[H]
\centering
\includegraphics*[width=0.5\textwidth]{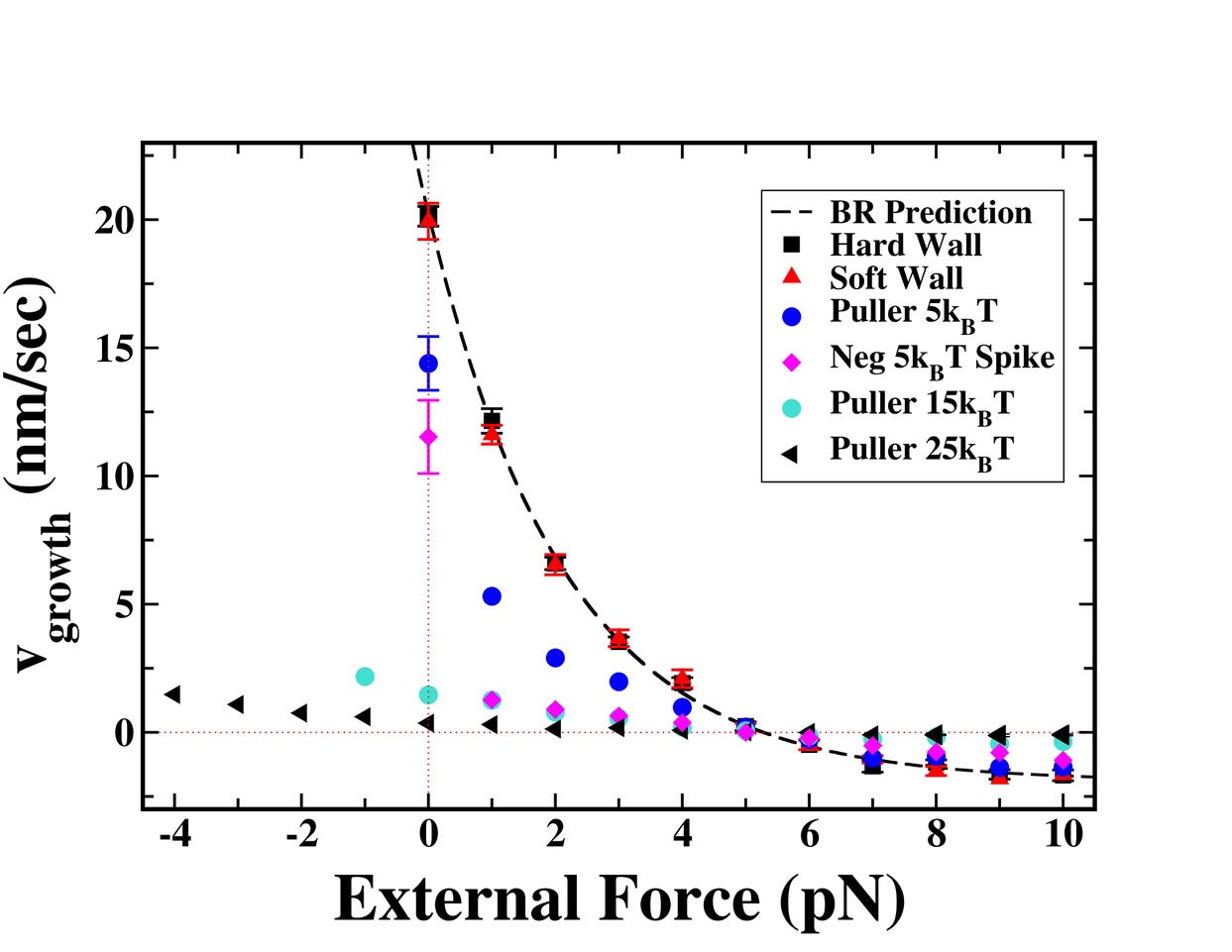}
\caption{Simulations of 20 seconds showing the force-velocity relation for different forms of filament-obstacle interaction potential for oblique filament orientation, including
filament-tip fluctuations. Error bars are smaller than the symbols, except where indicated. The dashed line is the prediction of the BR model at large diffusion constant values for obstacle. Data points for pullers are shown out to forces where they detach from the obstacle in less than 20\,sec.}
\label{O_F_V_plot}
\end{figure}

Figure \ref{O_F_V_plot} shows the resulting force-velocity relation for different pusher and puller potentials described in Figure \ref{Potentials}. 
Notice that the stall force is larger in this case because $\delta$ in Eq.\,\ref{fstall_formula} is replaced by $\delta \cos{(\theta)} = \delta/\sqrt{2}$. 
The BR relation remains  an upper bound for the growth velocity, and the force-velocity relations for both the hard wall and soft walls are very similar
to the BR relation. The $5k_BT$ well and spike potentials continue to have a zero-force velocity that is a substantial fraction
of the free-filament velocity, with velocities decaying more rapidly than the BR prediction.
In potentials with deep wells, the polymerization is slowed by roughly the same amount as in our baseline results (Figure \ref{F_V}).
The tradeoff between polymerization rate and maximum sustainable pulling force is also preserved. 

\section{Effect of Slow Obstacle Diffusion}
\setcounter{figure}{0}

\begin{figure}[H]
\centering
\includegraphics[width=0.45\textwidth]{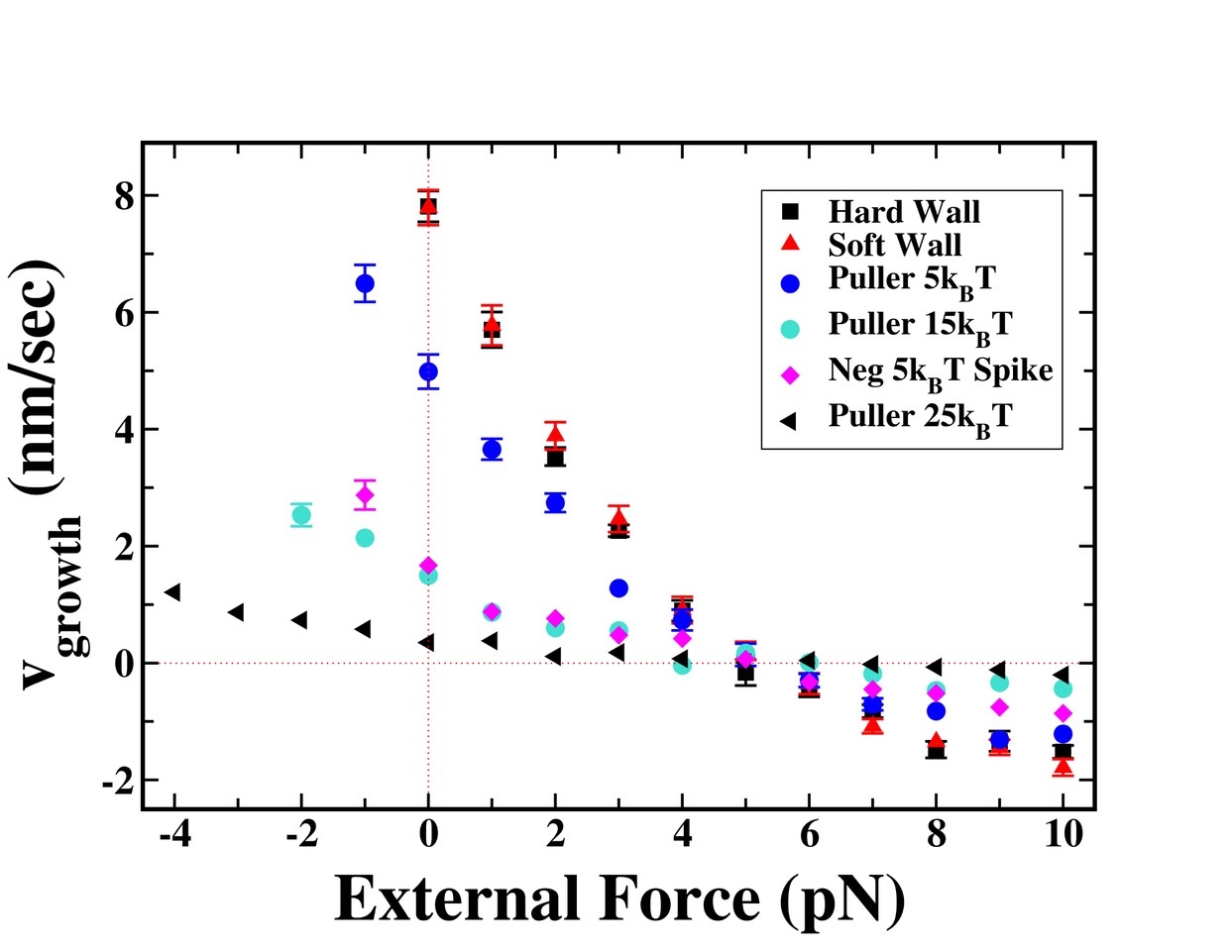}
\caption{Effect of reduced diffusion coefficient on force-velocity relation of growing actin filaments. 
Diffusion coefficient of $D$ =21 $nm/s^2$.  Force field parameters are as in Figure \ref{F_V}.}
\label{SlowDiffusion}
\end{figure}

We have repeated the force-velocity relations for the oblique filament case (45$^{\circ}$ orientation,
including filament-tip fluctuations) with a smaller diffusion coefficient to
see which findings in the main text depend strongly on the assumption of rapid diffusion. 
Figure \ref{SlowDiffusion} shows results for a diffusion coefficient of $D$ =21 $nm/s^2$. For this value of $D$, the dimensionless
parameter characterizing diffusion $\delta^2 k_{on}/2D_{obst}$ has the value unity, so the effects of diffusion
should be substantial. The polymerization rate for the more rapidly growing potentials - including the ``Hard Wall", the
``Soft Wall", and the ``Puller" with 5 $k_BT$ well depth, is slowed by about a factor of 2. As in the main text, the completely
repulsive continuous  potentials have force-velocity relations similar to the hard wall. 
However, the sharp drop in velocity for the $5k_BT$ potentials is eliminated,
as expected from the analysis of Sec. 3.2 showing that this effect depends on rapid diffusion. 
The effect on the other force-velocity relations for deeper wells is smaller.  
The general shape of the force-velocity is unchanged, and there is no effect on the
ordering of the curves. 

As Figure \ref{SlowDiffusion} indicates, the effect of using a softer wall on the growth velocity is minimal even when
diffusion is slow. For a still softer wall with decay coefficient $\kappa_1 = 0.2$ nm$^{-1}$ (data not shown), the acceleration is about
10\%. 

\end{document}